\documentclass[amssymb,amsmath,floatfix,twocolumn,superscriptaddress]{revtex4}
\usepackage{graphicx}
\usepackage{lastpage}
\usepackage{amssymb}
\newcommand{\ket}[1]{|#1 \rangle}

\begin{document}

\title{Integration of highly probabilistic sources into optical quantum architectures: perpetual quantum computation.}
    \author{Simon J. Devitt}
    \email{devitt@nii.ac.jp}
    \address{National Institute for Informatics, 2-1-2 Hitotsubashi, Chiyoda-ku, Tokyo 101-8430, Japan.}
 \author{Ashley M. Stephens}
 \address{National Institute for Informatics, 2-1-2 Hitotsubashi, Chiyoda-ku, Tokyo 101-8430, Japan.} 
      \author{William J. Munro}
    \address{NTT Basic Research Laboratories, NTT Corporation, 3-1 Morinosato-Wakamiya, Atsugi, Kanagawa 243-0198, Japan}
    \address{National Institute for Informatics, 2-1-2 Hitotsubashi, Chiyoda-ku, Tokyo 101-8430, Japan.}
        \author{Kae Nemoto}
    \address{National Institute for Informatics, 2-1-2 Hitotsubashi, Chiyoda-ku, Tokyo 101-8430, Japan.}
\date{\today}

\begin{abstract}
In this paper we introduce a design for an optical topological cluster state computer constructed 
exclusively from a single quantum component.  Unlike previous efforts we eliminate the need for 
on demand, high fidelity photon sources and detectors and replace them with the same device 
utilised to create photon/photon entanglement.  
This introduces highly probabilistic elements into the optical architecture while maintaining complete specificity 
of the structure and operation for a large scale computer.  
Photons in this system are continually recycled back into the preparation network, allowing 
for a arbitrarily deep 3D cluster to be prepared using a comparatively small number of 
photonic qubits and consequently the elimination of high frequency, deterministic photon sources.  
\end{abstract}
 
\maketitle
\section{Introduction}
Designing and constructing a viable large scale quantum information processing system 
has been the focus of extensive research ever since the first basic architectures were proposed in the 
1990's~\cite{CZ95,CFH97,GC97,K98,LD98,KLM01,MOLTW99,NPT99,ARPA,KMNRDM07}.  Experimental 
progress towards this goal has been 
pronounced~\cite{CNHM03,YPANT03,CLW04,HHB05,GHW05,G06,H06,G07,OPWRB03} and 
researchers are becoming slowly optimistic about the future potential for building a large scale computer. 

In recent years, the extraordinary advances in experimental systems and theoretical techniques for quantum 
information processing has allowed for serious questions in architectural design and construction to be 
discussed~\cite{DFSG08,HFJR10,MLFY10,DFTMN10,JMFMKLY10,DMN08,FWHLMH10,MHSDN10}.  
This new area of research is generally being referred to as Quantum Engineering (QE).  
Broadly, the primary goal of QE is to adapt and combine the best 
experimental technologies and theoretical techniques to construct an experimentally viable 
large-scale quantum computer.

One proposed architecture was introduced in 2009~\cite{DFSG08}.  This optics based computer is based on the 
topological cluster state model of computation~\cite{RH07,RHG07,FG08} and a photon-atom-photon coupling 
device called the photonic module~\cite{DGOH07}.  This architecture illustrated the structure and operation of a fault-tolerant, 
fully error corrected quantum architecture.  However, the architecture was based on components which were {\em all} theoretically 
deterministic.  
Deterministic photon-photon coupling utilized the non-linearity afforded by an atom/cavity system 
present within each photonic module and photonic sources and detectors were simply assumed to be deterministic and 
of high fidelity.

It has been well known since the seminal paper of Knill, Laflamme and Milburn~\cite{KLM01} that an all 
optical quantum computer based on linear elements only allows for coupling between qubits in a probabilistic 
fashion~\cite{SNMK03,SMENK06}.  
Since this result there has been extensive work investigating how probabilistic techniques could be utilized to 
construct a viable architecture~\cite{YR03,N04,HGMR04,BR05+,ND05,DR05,DHN06,GKE06,GHR07,KRE07,KGE07,KMNRDM07}.  
However, while this work demonstrated that in principal probabilistic components could be used to slowly grow large 
entangled states suitable for quantum computation two problems remained.  The first is that two dimensional cluster 
states~\cite{RB01} do not include any protocols for quantum error correction.  This problem was addressed 
by applying error correction protocols on top of the underlying cluster model and resulted in extremely high demands on 
quantum resources~\cite{DHN06}.  The second significant problem is that these results did not show how such a 
massive optical system was to be constructed, arranged and operated when large scale algorithms and 
error correction require billions of photons and millions of optical components.  

The experimental development of optical systems~\cite{OPWRB03,GPWRZ04,O07,LBYP07,LGZZYP08,L09,OFV09,PMO09} 
(and more generally, probabilistic quantum components~\cite{MMOY07,OHMMMM10+}) has in many ways been far more 
successful than deterministic technologies~\cite{MMKIW95,CPHS98,SKC00,CLW04,HHB05,SPS07,MCGK07}.  
Therefore, an important problem is can these more advanced probabilistic technologies 
still be integrated and used in viable large scale quantum information architectures.  The majority of research attempts to 
do this from the bottom up, taking well established non-deterministic protocols and incorporating 
them into appropriate computational and error correction models~\cite{LBSB10,FT10}.  While this addresses some issues related to more 
effective use of error correction techniques, these results still do not address the fundamental architectural structure 
of a large scale system.

We will attempt to approach this from the top down.  The optical architecture proposed in Ref.~\cite{DFSG08} 
is designed such that its physical structure and operation can be very well defined up to billions of qubits.  
Since the architectural structure of this system is so well defined, this will be our starting point.  The 
system is constructed from three key deterministic components.  We will show how we can remove two of them.  
We will take the optical architecture consisting of deterministic single photon sources, 
deterministic coupling via the photonic module and deterministic single photon detection and replace the 
sources and detectors with the photonic module.  This will lead to a network running with 
highly probabilistic single photon sources and entirely constructed from 
one quantum component, namely the photonic module.  

To combat the issue of probabilistic sources we introduce a perpetual design.  As the photonic module can 
act as a non-demolition photon detector, photons are simply recycled.  In this way, probabilistic sources are 
responsible for two tasks.  (1) Providing the photons to initialize the network and (2) replacing photons which 
are periodically lost during computation.  This paper will demonstrate how a highly probabilistic 
source can be integrated into a large scale architecture without sacrificing performance or the overall 
design and operation of the system.  This represents the first step at integrating probabilistic technologies 
with deterministic technologies and will hopefully lead to future architectural designs incorporating an increasing amount 
of probabilistic components.

We begin in section~\ref{sec:topological} with a brief discussion of the general principals governing the
topological cluster state model.  In section~\ref{sec:computer} we review the optical computer introduced in 
Ref.~\cite{DFSG08}.  Section~\ref{sec:reduction} discusses how we can reduce the required quantum technologies by replacing 
single photon sources and detectors with the same device utilized in the preparation of the cluster. 
Section~\ref{sec:perpetual} introduces the idea of a perpetual architecture design, where a comparatively small 
number of photons are recycled again and again to perform large cluster computations.  Each element of the architecture is 
detailed, including how fault-tolerance is maintained and how photon loss is reliably detected.  
The network consisting of photonic sources is discussed in Section~\ref{sec:injection}, where highly probabilistic sources 
are used to ``boot-up" the computer and replace heralded loss events during computation.  Finally in 
sections.~\ref{sec:simulations} and~\ref{sec:results} we present several network simulations illustrating that 
a perpetual computer utilizing photon recycling and highly probabilistic sources can operate effectively.

\section{Topological Cluster State Computing}
\label{sec:topological}
Topological cluster states were introduced by Raussendorf, Harrington and Goyal in 
2007~\cite{RH07,RHG07}.  This model incorporates the ideas from 
topological quantum computing~\cite{K97} and cluster state computation~\cite{RB01}.  This model 
for quantum information processing is very attractive 
as it incorporates fault-tolerant quantum error correction by construction and exhibits a high 
fault-tolerant threshold.  

\begin{figure}[ht!]
\begin{center}
\resizebox{55mm}{!}{\includegraphics{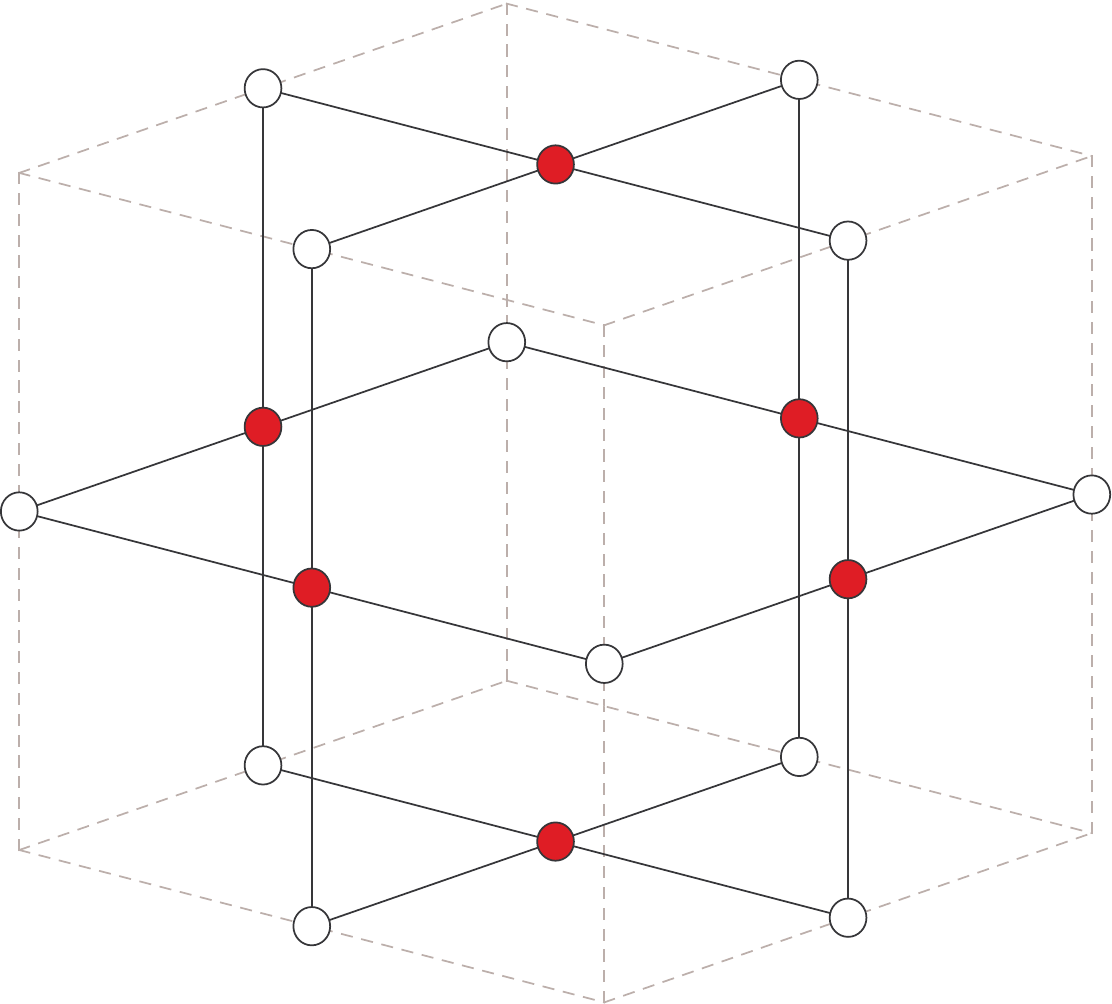}}
\end{center}
\vspace*{-10pt}
\caption{Unit cell of the 3D cluster required for topological cluster computing.  Each node represents a physical qubit initialized in the 
$\ket{+} = (\ket{0}+\ket{1})/\sqrt{2}$ state and each edge is a controlled-$\sigma_z$ operation between two qubits.  
The 2D cross section of the cluster dictates the size (and 
error correcting power) of the computer while the third dimension dictates the total number of 
computational time steps available.  Computation proceeds by measuring the cluster along one of the three dimensions.}
\label{fig:cell}
\end{figure}

\begin{figure}[ht!]
\begin{center}
\resizebox{85mm}{!}{\includegraphics{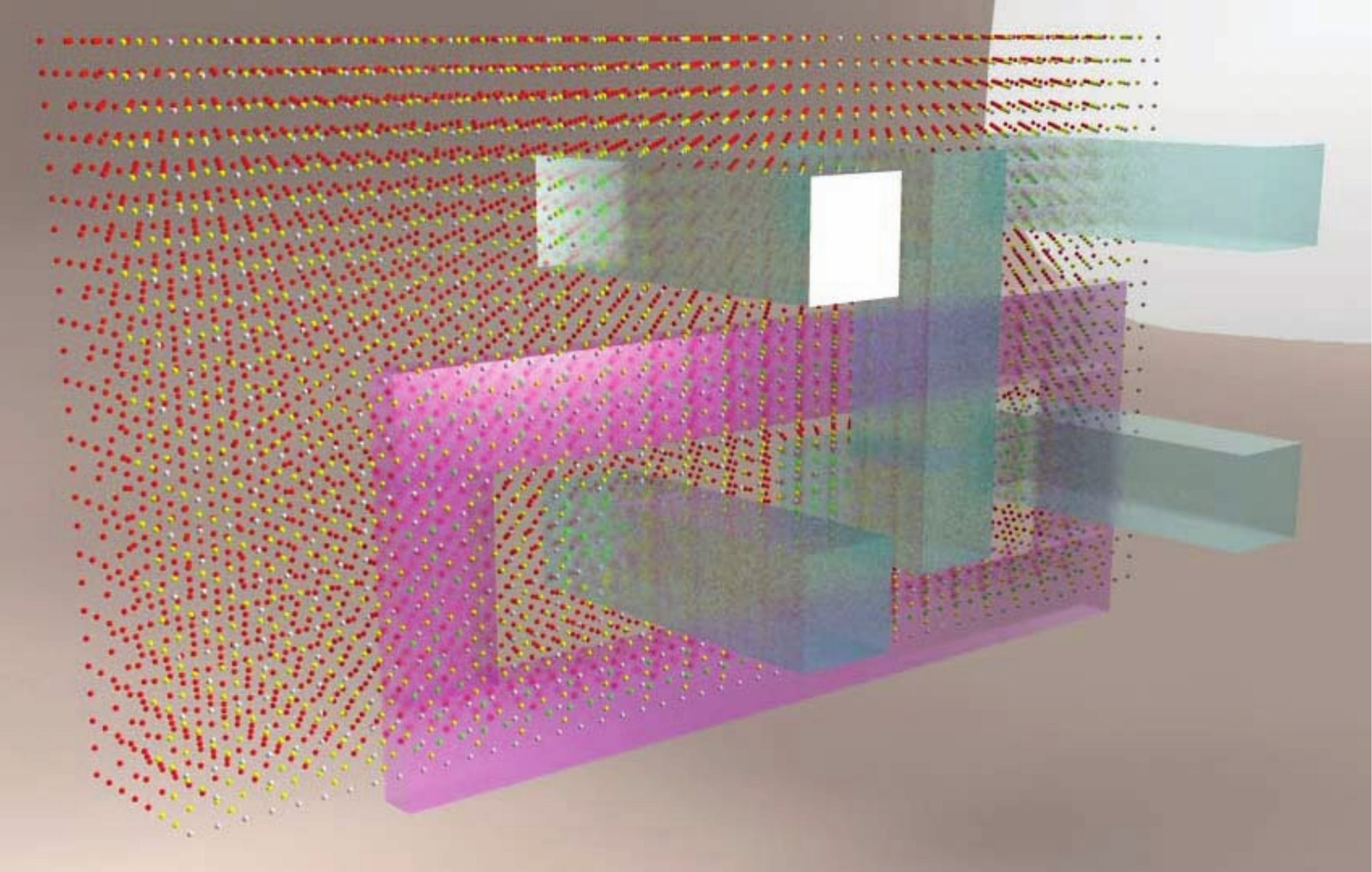}}
\end{center}
\vspace*{-10pt}
\caption{Diagram of a logical CNOT operation in the topological cluster model.  
Each point in the diagram represents a physical qubit and 
the cluster is consumed from the front of the image to the rear.  This image shows a CNOT which is 
approximately 50\% complete.  Logical 
qubits are defined via pairs of defects.  Four sets of defects are required to perform the braided CNOT.  The three blue defects represent the 
control input, control output and target respectively.  
These three defects are known as primal defects~\cite{RHG07,FG08}.   
The purple defect is known 
as a dual defect and is used to enact braided logic operations (logic operations can only be performed between defects of differing type).  
The total size and separation of each defects dictates the error correcting power of the topological code.}
\label{fig:CNOT}
\end{figure}

Computation proceeds via the initial construction of a highly entangled multi-qubit state.   Fig.~\ref{fig:cell} illustrates 
a unit cell of the cluster.  Each node represents a physical qubit, initially prepared in 
the $\ket{+} = (\ket{0}+\ket{1})/\sqrt{2}$ state, and each edge represents a controlled-$\sigma_z$ 
entangling gate between qubits.  This unit cell of the cluster extends in all three dimensions, dictating the size and 
error correcting strength of the computer.  
Computation under this model is achieved via the consumption of the cluster along one of the 
three spatial dimensions~\cite{RHG07,FG08} (referred to as simulated time).  Logical
qubits are defined via the creation of ``holes" or ``defects" 
within the global lattice and multi-qubit operations are achieved via braiding operations 
(movement of defects around one another) as the cluster is consumed along the direction 
of simulated time.  Fig.~\ref{fig:CNOT} illustrates a braided CNOT operation.  Qubits within the 
cluster are selectively measured in the $\sigma_z$ basis to create and manipulate defects.  By measuring the correct 
set of physical qubits, defects can be moved as the cluster is consumed.  Physical qubits not associated with defects are 
measured in the $\sigma_x$ basis and are utilized to perform fault-tolerant error correction on the system.

The specific details for computation under this model are not important for this 
discussion and we encourage the reader to refer to Refs.~\cite{RHG07,FG08} for further details.  
For this analysis, we will be examining the network required to successfully create the entangled cluster for computation.

\section{Optical Topological Computer}
\label{sec:computer}
\begin{figure}[ht!]
\begin{center}
\resizebox{85mm}{!}{\includegraphics{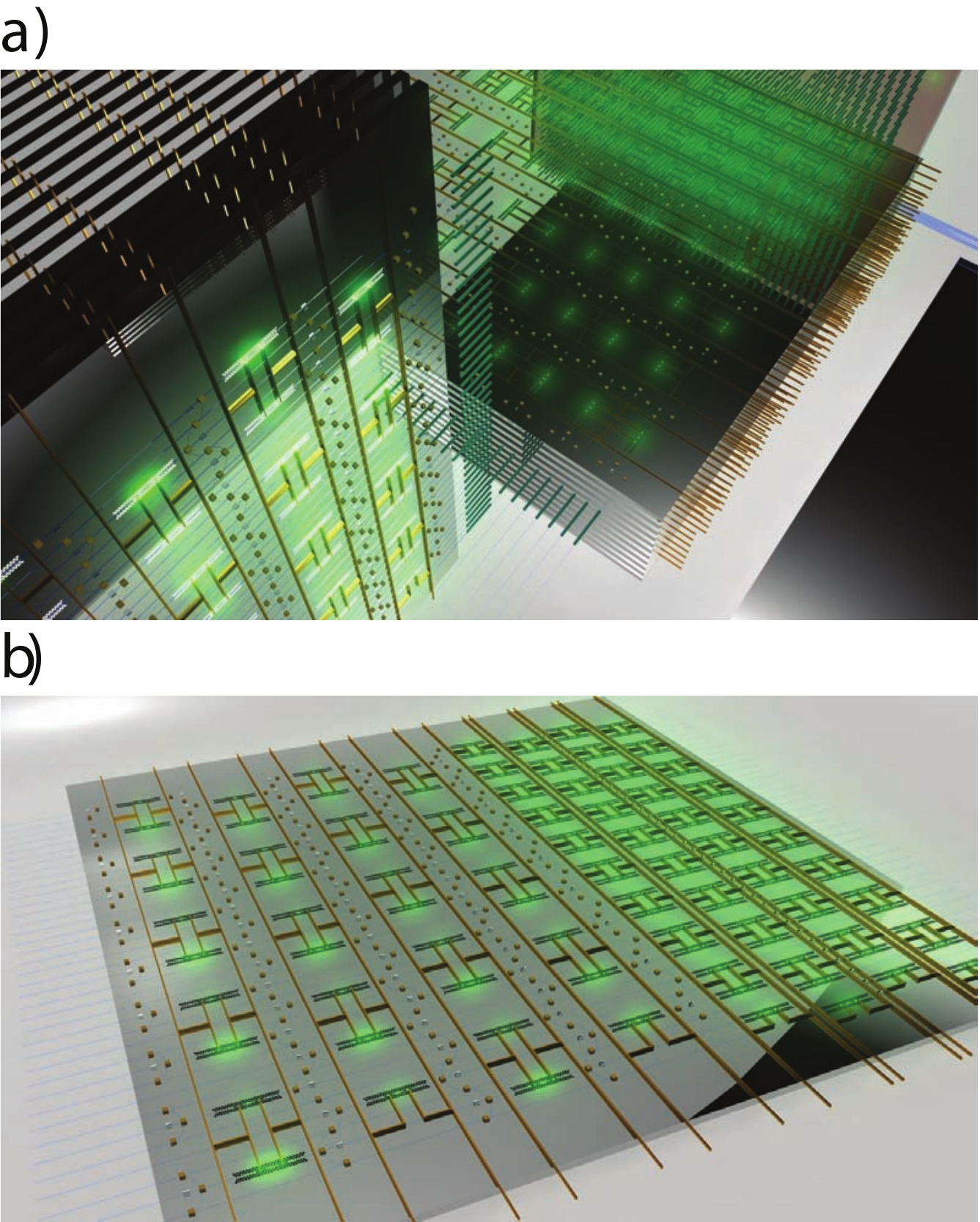}}
\end{center}
\vspace*{-10pt}
\caption{Structure of the optical topological computer introduced in Ref.~\cite{DFSG08}.  The optical computer is constructed via a stacked array 
of ``wafers".  Each wafer consists of a network of photonic modules~\cite{DGOH07}, each of which is designed to deterministically couple 
single photons.  Subfigure a) illustrates the structure of the actual computer.  two sets of wafters, oriented at 90$^\circ$ to each other are 
connected to prepare the required entangled cluster.  Photons travel through the network from left to right through 
four stages of cluster preparation before they are measured.  Subfigure b) illustrates stages three and four of the cluster preparation 
network and a second array of photonic modules used for measurement [Sec.~\ref{sec:measurement}].  The width of each wafer is 
independent of the total size of the computer while the length increases linearly with the size and/or error correcting power of the 
computer.}
\label{fig:computer}
\end{figure}

Shown in Fig.~\ref{fig:computer} is the basic structure of the 
preparation network for the computer.  The preparation 
network consists of two sets of fabricated ``wafers" containing an interlaced network of 
photonic modules~\cite{DGOH07} oriented at 90$^\circ$ to each other [Fig.~\ref{fig:computer}a)].  Each 
set of parallel wafers are {\em not} interconnected, the only connections are at the 
junction between horizontally and vertically oriented wafters.  In total there are four separate stages in the preparation network, two on 
each horizontal wafer and two on each vertical wafer.  The computer operates via the injection of single photons into the left hand 
side of this network.  Each photon then interacts with a total of four individual photonic modules which acts to deterministically 
entangle photons into the required cluster.  
After photons are entangled, they are then measured in appropriate bases to perform fault-tolerant, error corrected quantum computation. 

Shown in Fig.~\ref{fig:computer}b) is a schematic of one of the wafers.  This wafer illustrates stages three and four of the cluster 
preparation network and an additional array of photonic modules that are utilized to perform measurement of the cluster 
(this will be discussed in Sec.~\ref{sec:measurement}).  The total width of each of these wafters remains constant, while the length of 
each of these wafters scales linearly with the size of the computer.  

\begin{figure*}[ht!]
\begin{center}
\resizebox{150mm}{!}{\includegraphics{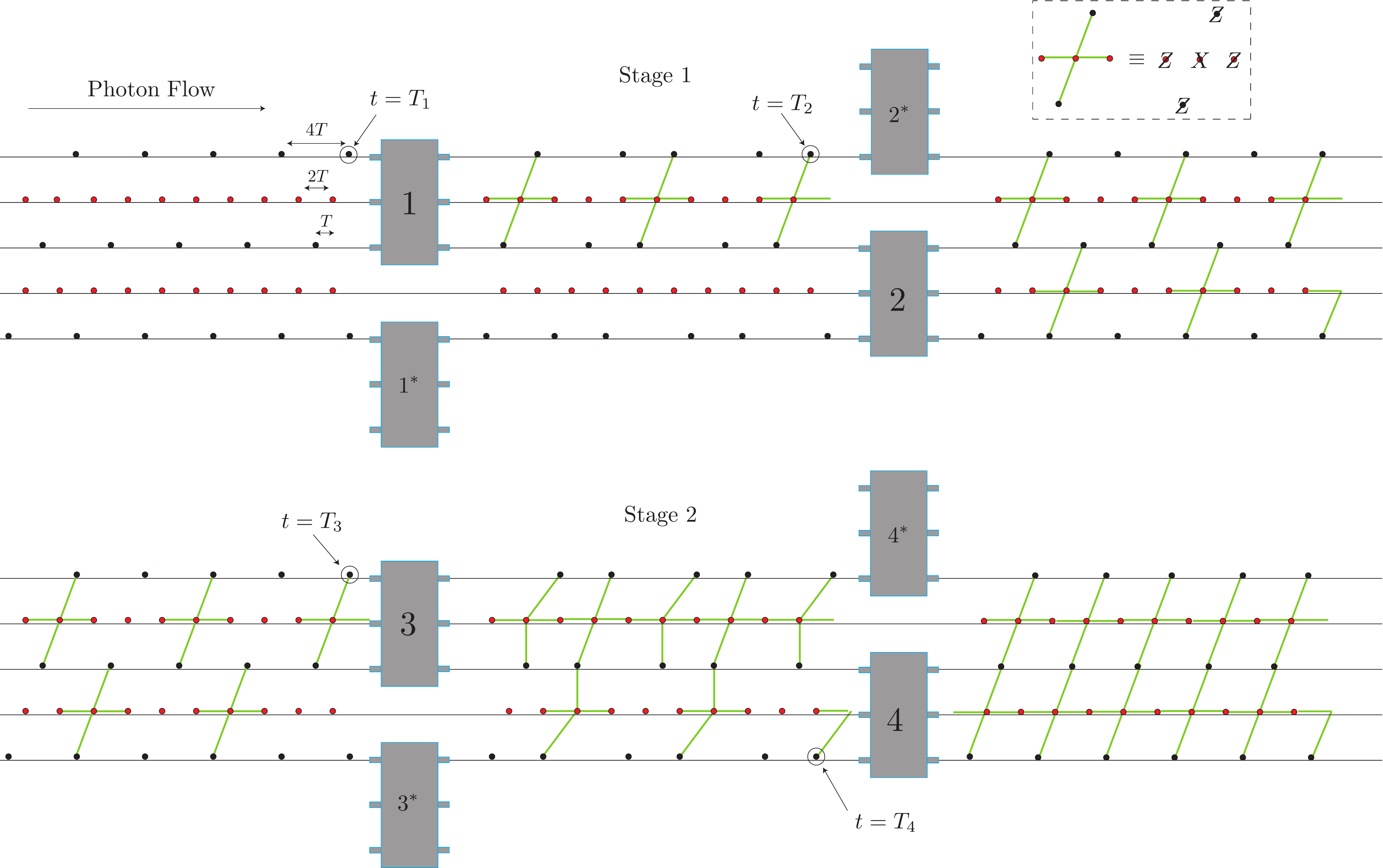}}
\end{center}
\vspace*{-10pt}
\caption{Temporal arrangement for stages one and two of the cluster preparation network.  Half of the optical lines have 
photons separated by $4T$ while the other half of the optical lines have photons separated by $2T$, where $T$ is 
the operational time for a photonic module~\cite{DGOH07}.  This temporal arrangement allows for each photon to 
flow through the network of modules.  After exiting the network, each photon is immediately measured to 
perform computation.}
\label{fig:sequence}
\end{figure*}

The injection of single photons into the computer requires a specific temporal arrangement.  Illustrated in Fig.~\ref{fig:sequence} 
is the temporal arrangement of photons for stages one and two of the preparation network 
for one of the vertically oriented wafers shown in Fig.~\ref{fig:computer}a).  Within the network, half of the 
optical lines contain photons temporally separated by $2T$ and the other half have photons temporally separated by 
$4T$, where $T$ is the operational time of a single photonic module.  This arrangement allows the photon stream to essentially ``flow" 
through the interlaced network of modules, maintaining temporal separation and ensuring that only a single photon is 
present within any photonic module at any given time.  Utilizing this temporal arrangement, it was shown in Ref.~\cite{DFSG08} that this 
network could deterministically prepare an arbitrarily large 3D topological cluster without employing sophisticated photonic routing or 
storage.  Additionally the specific operation of {\em every} photonic module in the network is completely specified and independent of the 
size of the computer.  

\section{Reducing required technologies}
\label{sec:reduction}
The optical computer requires several different, high fidelity, quantum components. 
(1) The Photonic Module~\cite{DGOH07}.
(2) High frequency, on demand, single photon sources.
(3) High fidelity single photon detectors.
(4) High fidelity single photon switching.
It could be argued that the construction of large numbers of photonic modules represents the most difficult element of the 
above list.  If the construction of such a device could be 
achieved, it could be safely assumed that the other required quantum components will also exist 
with sufficient reliability to construct a large scale computer.  However, what if this is not the case? Can we further reduce the 
required technology and still design a viable architecture?  

Aside from the photonic module, the next two components listed above that will require extensive development are 
high frequency, on demand, single photon sources and high fidelity single photon detectors.  These components currently exist 
at several levels of efficiency and reliability~\cite{KMNRDM07,SSRG09}.  However, these technologies still 
require significant work before they are adaptable to a large scale 
quantum computer.  These are the two components that we will eliminate.  We will replace high fidelity single photon 
detectors with the photonic module itself and we will replace the high frequency, on demand, sources with non-deterministic 
sources.  While we could consider multiple non-deterministic sources, we will assume that photonic modules 
are used, distilling weak coherent light into single photons with a very low probability of success.  
In this way we will redesign the computer architecture to 
consist of only two required components, a high fidelity photonic module and reliable single photon switching.  

It should be stressed that while the goal of this redesign may seem theoretically trivial, from an architectural standpoint it is 
significantly more complicated.  In principal, any number of techniques could be utilized to remove the requirements of 
single photon sources and detectors.  However, we need to accomplish this under some tight constraints.  
\begin{enumerate}
\item Original design of the architecture required one high frequency (MHz or above) source per optical line or an ultrahigh frequency 
source (THz or above) serving multiple optical lines.  When replacing deterministic sources with non-deterministic sources, this ratio of 
the number of sources per optical line must remain effectively constant.  
\item Non-deterministic sources such as the photonic module 
prepare single photons at random (but heralded) times.  As these photons may be required anywhere 
in the network, the design must allow us for this without introducing complicated photon routing or significant photon storage.
\item The probability of photon loss {\em within} the computer is non-zero.  Therefore the design must be able to effectively replace 
lost photons, which will occur in random but heralded locations, with new photons which will also be prepared 
at random but heralded locations.
\end{enumerate}

\subsection{Utilizing the photonic module as a detector}
\label{sec:detectors}
The ability to utilize the photonic module to perform individual photon measurement is the key to replacing on demand 
sources without reintroducing additional technologies.  As described in Ref.~\cite{DGOH07} the sole action of the photonic module is 
to project an arbitrary $N$-photon state, $\ket{\psi}_N$ into a $\pm 1$ eigenstate of the operator $X^{\otimes N}$.

Within the preparation network of the computer, each module interacts with five photons before being measured 
and decoupled.  This, combined with suitable local operations, projects each five photon group into $\pm 1$ eigenstates 
of the operator $ZZXZZ$.  These are the relevant eigen-operators describing the topological cluster state utilized for 
computation~\cite{RH07,RHG07,DFSG08}.   
If, however, only a single photon is allowed to interact with a photonic module between its initialization and measurement, an 
arbitrary state $\ket{\psi}$ is mapped to,
\begin{equation}
M\ket{\psi}\ket{+}_a = \frac{1}{2}(\ket{\psi} + X\ket{\psi})\ket{g_1}_a + \frac{1}{2}(\ket{\psi}-X\ket{\psi})\ket{g_2}_a
\end{equation}
where $M$ is the interaction between the atomic system in the module and the photon, $\ket{g_1}_a$, $\ket{g_2}_a$ are 
the two states of the atomic qubit and $\ket{+}_a = (\ket{g_1}_a+\ket{g_2}_a)/\sqrt{2}$.  
Measuring the atomic system will project the photon into either 
the $\ket{+} = (\ket{0}+\ket{1})/\sqrt{2}$ or $\ket{-} = (\ket{0}-\ket{1})/\sqrt{2}$ state dependent on measuring the module in 
the $\ket{g_1}_a$ or $\ket{g_2}_a$ state respectively.  Therefore, the module can be utilized to perform an $X$ basis measurement 
on any photon it is allowed to interact with between initialization and measurement.  
Combining this with appropriate local rotations before measurement (via optical 
waveplates or waveguide techniques~\cite{O07}) allows for the measurement of a single photon in any desired basis.  

The major advantage to using the module is that it is intrinsically a non-demolition measurement.  The photonic 
state is measured via readout of the {\em atomic} system.  Therefore the photon is not physically destroyed during measurement 
and will simply exit the module in exactly the same way as for the preparation network.  Consequently, we can recycle it.  Measured 
photons can therefore be rerouted back into the input of the cluster preparation network.   

\subsection{Using the photonic module as a probabilistic source}
\label{sec:sources}
By using photon recycling we can, in principal, operate a large topological cluster state computer with a comparatively small number of 
individual photons.  However, these photons still need to be initially prepared and loss events within the network need to be compensated 
with some type of source device.  

In Ref.~\cite{DFSG08} it was assumed that the cluster network was fed via high frequency, on demand, single 
photon sources.  Assuming one such source per optical line in the network, the operational frequency of each source between 500KHz 
and 100MHz (assuming that the photonic modules have an operational time between 10ns and 1$\mu s$~\cite{DGOH07,SGMNH08}).
Replacing a high frequency source with a probabilistic source and photon recycling can, in many circumstances, be 
desirable.  In the case of our architecture, the photonic module can act as this probabilistic source.  This allows for a large scale 
quantum architecture to be constructed using essentially only one quantum component.  

The photonic module acts as a parity check device, allowing us to effectively measure multi-qubit observables.  It is well known that such 
a quantum component can be utilized for distilling weak coherent light into single photon states~\cite{NM05,MNS05}.  
A weak coherent pulse, in the number basis, can be approximated as,
\begin{equation}
\ket{\alpha} \approx e^{\frac{-|\alpha|^2}{2}} \left( \ket{0} + \alpha\ket{1} + \frac{\alpha^2}{\sqrt{2}}\ket{2} + \frac{\alpha^3}{\sqrt{6}}\ket{3} \right)
\end{equation}
for $\alpha \ll 1$.  This state can routinely be prepared in the laboratory.  

\begin{figure}[ht!]
\begin{center}
\resizebox{50mm}{!}{\includegraphics{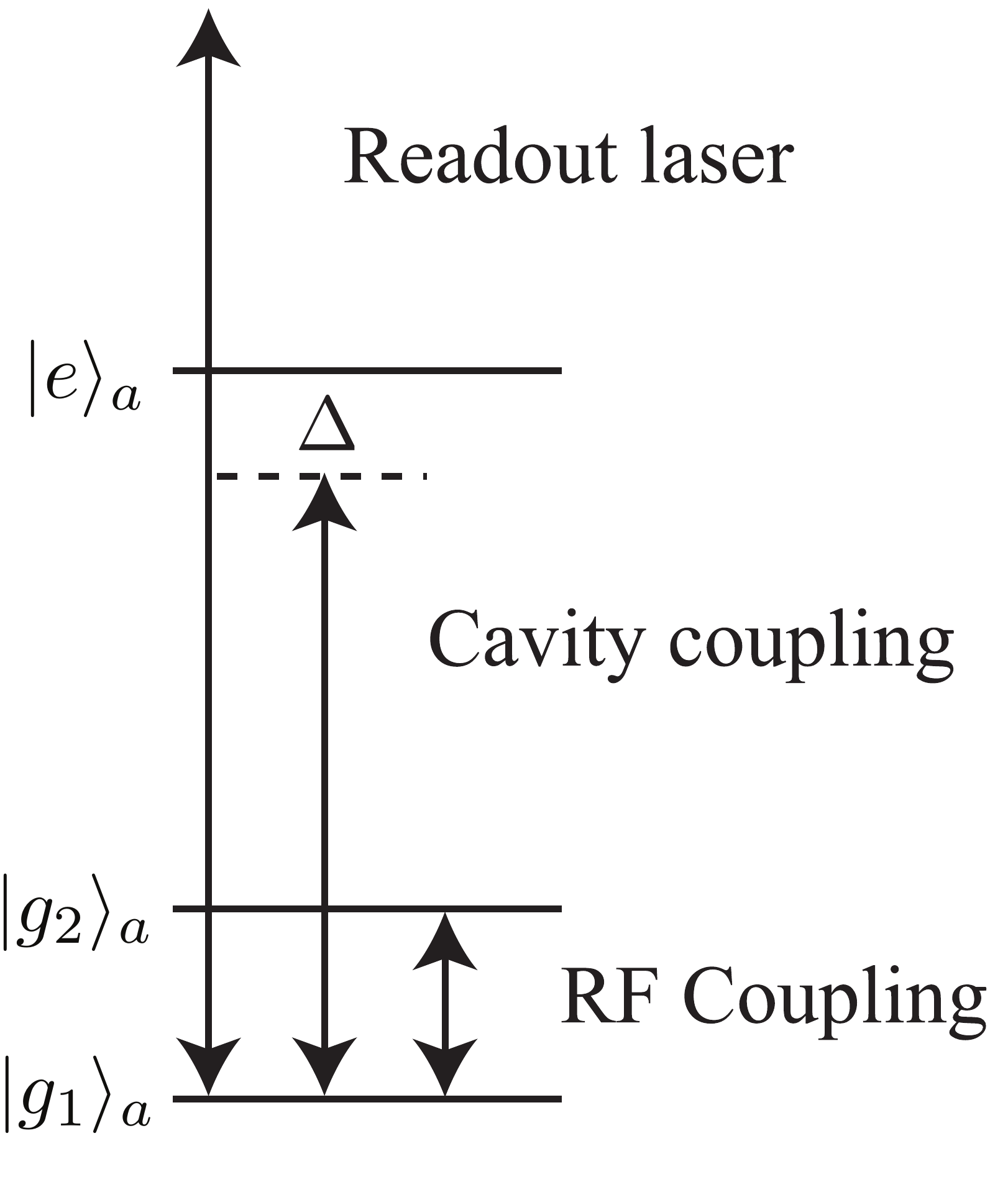}}
\end{center}
\vspace*{-10pt}
\caption{Internal structure of the atomic system utilized in the photonic module.  The two ground states, $\ket{g_1}_a$ and $\ket{g_2}_a$ 
are the two levels measured when each module is read-out.  The transition between $\ket{g_1}_a$ and a 
third state $\ket{e}_a$ is detuned from the cavity mode 
by an amount $\Delta$.  Utilizing a basic Jaynes-Cummings interaction in the dispersive limit, a phase shift will accumulate on the 
state $\ket{g_1}_a$ if the cavity mode is occupied with a single photon.}
\label{fig:module}
\end{figure}
The internal mechanism of the photonic module is based on working in the dispersive limit of the Jaynes-Cummings model.  Illustrated in 
Fig.~\ref{fig:module} is the structure for the internal atomic system for each photonic module.  In the dispersive limit, the effective 
Hamiltonian of the system is, $H = \beta a^{\dagger}a \sigma_z$.  The operator(s) $a$ ($a^{\dagger}$) are the creation (annihilation) operators 
for the cavity mode, $\sigma_z$ is the usual Pauli operator over the qubit spanned by the two ground states $\ket{g_1}_a$ and 
$\ket{g_2}_a$ and the effective coupling constant is $\beta = -g^2/\Delta$.  
If the cavity mode is detuned by $\Delta$ 
from the transition $\ket{g_1}_a \leftrightarrow \ket{e}_a$ and the atomic system is initialized in the state $(\ket{g_1}_a+\ket{g_2}_a)/\sqrt{2}$, 
then a single photon introduced into the cavity mode will result in the following evolution,
\begin{equation}
\frac{1}{\sqrt{2}}\left( \ket{g_1}_a + \ket{g_2}_a\right )\ket{1} \rightarrow \frac{1}{\sqrt{2}}\left( e^{2i\beta t}\ket{g_1}_a + \ket{g_2}_a\right )\ket{1},
\end{equation}
for an interaction time $t$.  If we tune the interaction time to $t = \pi/(2\beta)$ before removing the photon from the cavity and 
out-coupling it to a waveguide, we will induce a $\pi$ phase shift on the state $\ket{g_1}_a$.  

Instead of a single photon we now interact the atomic system with a weak coherent pulse.  The system evolution is,
\begin{equation}
\begin{aligned}
\frac{1}{\sqrt{2}}\left (\ket{g_1}_a+\ket{g_2}_a\right) \ket{\alpha} \rightarrow 
&\frac{e^{\frac{-|\alpha|^2}{2}}}{\sqrt{2}}\left(\ket{g_1}_a+\ket{g_2}_a\right)\ket{0} 
\\
&+ \frac{\alpha e^{\frac{-|\alpha|^2}{2}}}{\sqrt{2}}\left(-\ket{g_1}_a+\ket{g_2}_a\right)\ket{1} \\
&+ \frac{\alpha^2 e^{\frac{-|\alpha|^2}{2}}}{2}\left(\ket{g_1}_a+\ket{g_2}_a\right)\ket{2} \\
&+ \frac{\alpha^3 e^{\frac{-|\alpha|^2}{2}}}{2\sqrt{3}}\left(-\ket{g_1}_a+\ket{g_2}_a\right)\ket{3}.
\end{aligned}
\label{eq:module2}
\end{equation}
Defining the states, $\ket{\pm}_a = (\ket{g_2}_a\pm \ket{g_1}_a)/\sqrt{2}$, and rewriting Eq.~\ref{eq:module2}, we have,
\begin{equation}
e^{\frac{-|\alpha|^2}{2}}\ket{+}_a(\ket{0}+\frac{\alpha^2}{\sqrt{2}}\ket{2}) + \alpha e^{\frac{-|\alpha|^2}{2}}\ket{-}_a(\ket{1}+\frac{\alpha^2}{\sqrt{6}}\ket{3})
\end{equation}
The atomic system is measured in the $\ket{\pm}_a$ basis.  
The probability of each outcome is~\footnote{We calculate the probabilities associated 
with each measurement outcome using the full coherent state $\ket{\alpha} = e^{-|\alpha|^2/2}\sum_{n=0}^{\infty}\frac{\alpha^n}{\sqrt{n!}}\ket{n}$, before making the approximation $\alpha \ll 1$.},
\begin{equation}
\begin{aligned}
P(\ket{+}_a) &= e^{-|\alpha|^2} \sum_{n=0}^{\infty} \left(\frac{|\alpha|^{4n}}{2n!}\right) \\
&= e^{-|\alpha|^2}\cosh (|\alpha|^2) \approx 1- |\alpha|^2,  \\
P(\ket{-}_a) &= e^{-|\alpha|^2} \sum_{n=0}^{\infty} \left(\frac{|\alpha|^{4n+2}}{(2n+1)!}\right) \\
&= e^{-|\alpha|^2}\sinh (|\alpha|^2) \approx |\alpha|^2, \\
\end{aligned}
\label{eq:prob}
\end{equation}
and the resultant states after each measurement result is,
\begin{equation}
\begin{aligned}
&\ket{\psi} \approx \sqrt{\frac{4}{4+|\alpha|^4}}\left(\ket{0}+\frac{\alpha^2}{\sqrt{2}}\ket{2}\right), \quad \text{if } \ket{+}_a \text{ is measured,} \\
&\ket{\psi} \approx \sqrt{\frac{6}{6+|\alpha|^4}}\left(\ket{1}+\frac{\alpha^2}{\sqrt{6}}\ket{3}\right), \quad \text{if } \ket{-}_a \text{ is measured.} 
\end{aligned}
\end{equation}
Therefore, if the atomic system within the module is measured in the $\ket{-}_a$ state, the projected optical state approximates a single photon.  
The strength of the weak coherent state is the determining factor in how close the projected eigenstate is to a single photon, with a 
fidelity given by,
\begin{equation}
F = |\langle1\ket{\psi}|^2 = \frac{6}{6+|\alpha|^4} \approx 1-\frac{|\alpha|^4}{6}.
\end{equation}
The relationship is inverted when considering the 
probability of measuring the module in the $\ket{-}_a$ state.  As $\alpha \rightarrow 0$, the probability 
of measuring the module in the $\ket{-}_a$ state and projecting the coherent state into an approximate single photon state approaches zero.  
Hence, there is a tradeoff between the probability that a module will successfully distill a single photon and 
how well the distilled state approximates a single photon.  

Utilizing the photonic module to distill weak coherent states will therefore result in sources with very low probabilities of success (as we 
require distilled states to approximate single photons to a high degree).  However, even with low success probabilities, modules combined 
with photon recycling can be combined successfully.

\section{Perpetual Network}
\label{sec:perpetual}
\begin{figure}[ht]
\begin{center}
\resizebox{80mm}{!}{\includegraphics{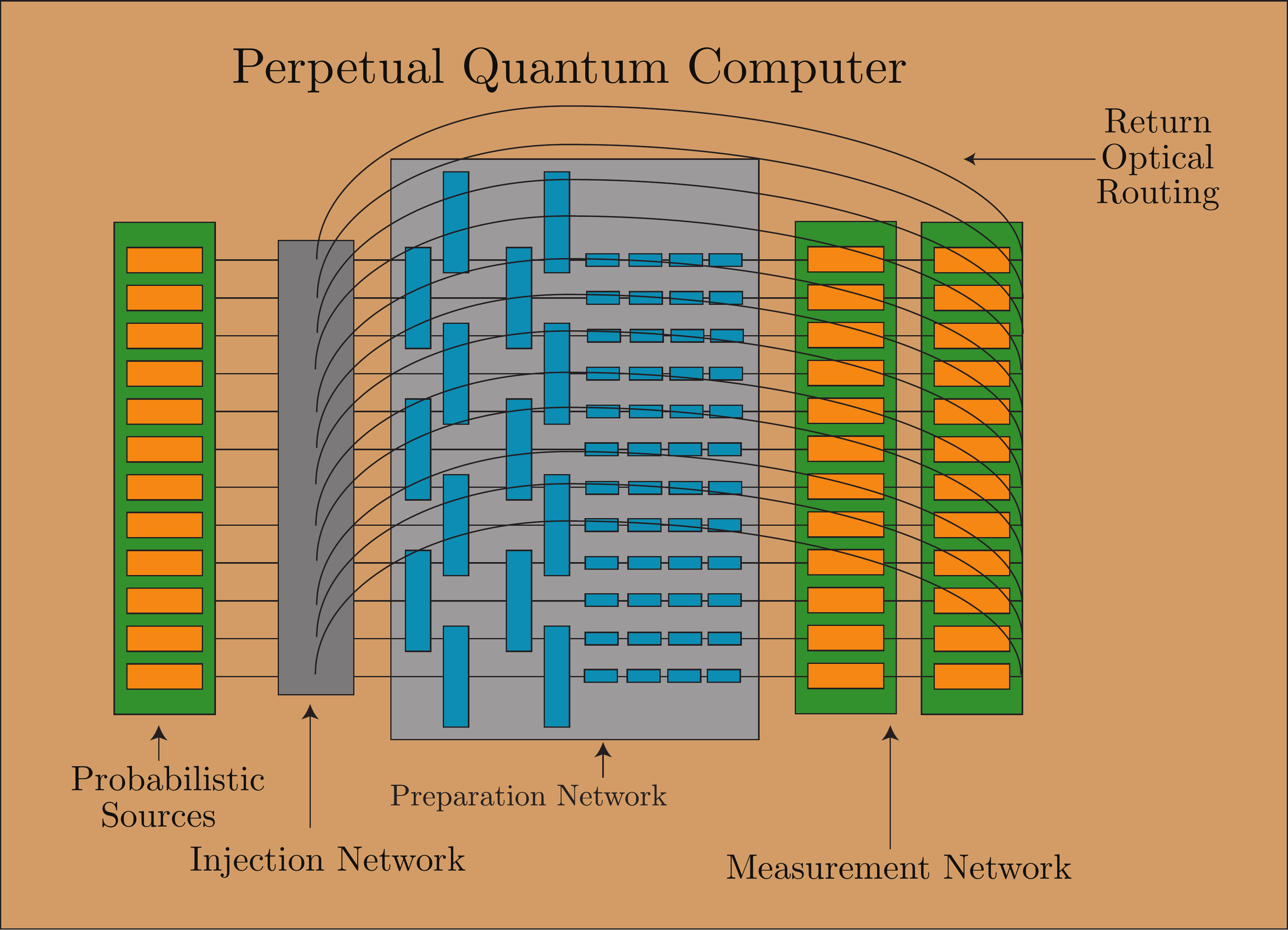}}
\end{center}
\vspace*{-10pt}
\caption{General structure of the preparation network.  Dedicated single photon sources and detectors are replaced with 
photonic modules.  As each module acts as a non-demolition detector, photons can be rerouted back into the 
start of the preparation network.  Each deterministic source can then be replaced with low probability sources which 
provides the initial photons to saturate the preparation network and replace photons lost during computation.}
\label{fig:network1}
\end{figure}

The previous section illustrated how the photonic module can be utilized to effectively replace single photon detection and probabilistically 
distill weak coherent states into single photons.  We can now discuss the general structure of a perpetual quantum computer.  
Fig.~\ref{fig:network1} illustrates the overall structure of the design.  
Probabilistic sources are used to slowly ``boot-up" the computer.  Each injected photon then proceeds through the preparation 
network of photonic modules, through a network of detection modules and then rerouted back to the source.  
We examine each component of the network separately,
\subsection{Preparation Network}
The preparation network has already been detailed in Ref.~\cite{DFSG08}.  The temporal arrangement of photons is such that half of the optical 
lines consist of photons separated by $2T$ and the other optical lines contain photons separated by $4T$, where $T$ is the 
operational time of the module.  Each photon interacts with four separate modules and suffers a delay of $T$ for each interaction.  

Within the network there are temporal ``windows" which allows for each module to be measured and reinitialized.  Both the measurement 
window and the reinitialization window are assumed to also take time $T$.  
After the preparation network, there is an additional delay of $4T$ before each photon enters the measurement network.  This delay allows 
the final parity checks in cluster preparation to be completed before photons are measured.

\subsection{Detection Network}
\label{sec:measurement}
As shown in Sec.~\ref{sec:detectors}, the modules can be utilized to perform non-demolition detection on each photon.  However, there is 
issues related to fault-tolerance that needs to be addressed when designing the detection system.  

It is well known that photon loss is a major error channel for optical computers.  The topological cluster codes that are used in this 
computer are quite efficient at correcting this type of error.  Several recent results have examined the robustness of the topological 
cluster model when subjected to loss~\cite{BS10}, 
suggesting that this particular error channel is preferable over standard quantum errors.  

\begin{figure}[ht!]
\begin{center}
\resizebox{45mm}{!}{\includegraphics{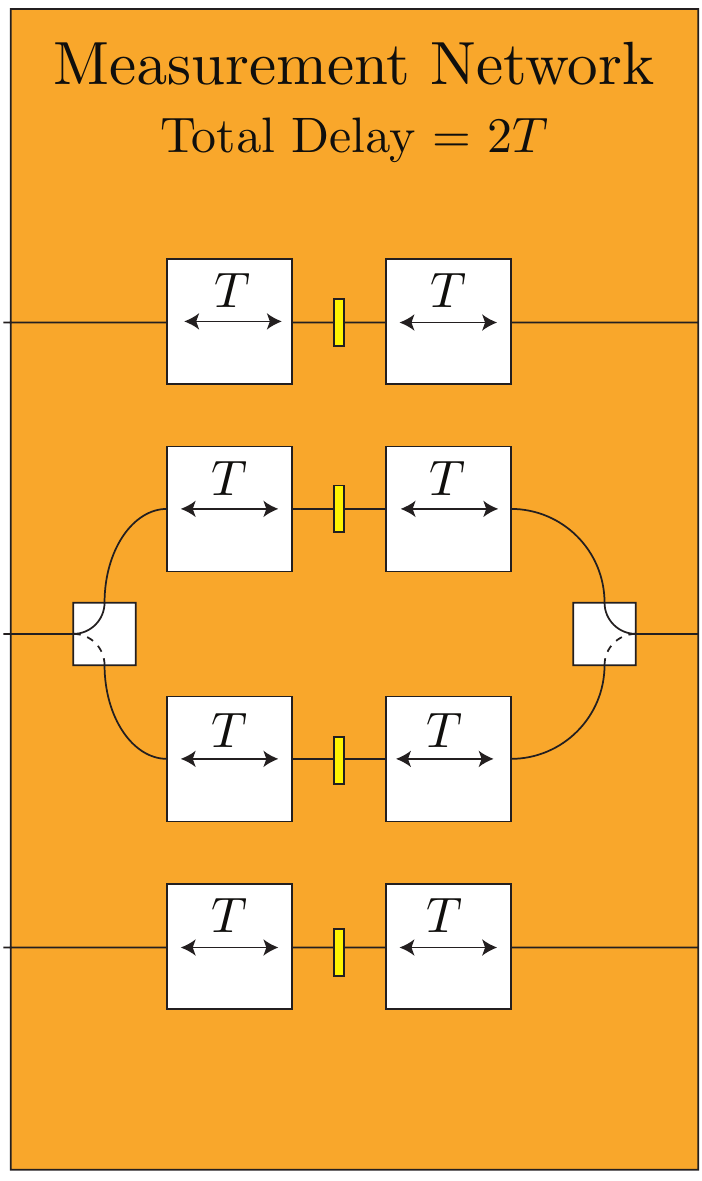}}
\end{center}
\vspace*{-10pt}
\caption{Detection network of photonic modules.  Illustrated is a small cross section of the network.  The upper and lower optical 
lines run at a temporal separation of $4T$ while the central line operates at a temporal separation of $2T$.  Loss can be uniquely identified 
by measuring photons twice separated by a single photon phase rotation.  As each module requires a temporal window of $T$ 
for measurement and reinitialization, the central optical line has two sets of measurement modules.  While one module is 
reinitialized, the next incident photon is routed to a second module that was initialized in the previous time step.}
\label{fig:detect}
\end{figure}

In the original design of the architecture, photon detection was achieved using dedicated single photon detectors which destructively 
measure photons and hence can discriminate between the presence or absence of the physical photon.  This combined with the 
fact that new photons are continuously injected from deterministic sources meant that loss was an easily correctable error for the 
original design.  However, moving to a perpetual architecture presents two problems.  
\begin{enumerate}
\item The internal function of the photonic module is such that it cannot discriminate between a photon in the $\ket{+}_a$ state and the 
vacuum. 
\item As photons are being recycled, undetected loss errors will be temporally correlated during computation (since the same physical photon 
is continuously re-entanged into the cluster at later times).
\end{enumerate}
Therefore, we need to first figure out a method to uniquely detect loss events using the modules and ensure that this technique can be 
made fault-tolerant (i.e. not cause errors in detection to spread to large groups of errors at later times).  
The detection  network is illustrated in Fig.~\ref{fig:detect}.

Illustrated is a small cross section of the network.  The upper and lower optical lines run at a repetition rate of $4T$ while the central 
optical line runs at a repetition of $2T$.  As we assume each photonic module requires a window of $T$ for both measurement and 
reinitialization, the detection network for the central optical line has twice as many modules.  While one of the modules is being reinitialized 
for the next measurement, the next incident photon arrives at the same time.  Therefore, it is switched to a second module 
which has already been initialized.  

Fault-tolerance is achieved via measuring photons twice.  We can write down the module transformations for an incident photon in the 
$\ket{+}$ state, the $\ket{-}$ state and the vacuum.  
\begin{equation}
\begin{aligned}
&M\ket{+}\ket{+}_a = \ket{+}\ket{+}_a\\
&M\ket{-}\ket{+}_a = \ket{-}\ket{-}_a\\
&M\ket{\text{vac}}\ket{+}_a = \ket{\text{vac}}\ket{+}_a.
\end{aligned}
\end{equation}
Hence, measuring the atomic system in the $\ket{+}_a$ state 
indicates the presence of either a single photon $\ket{+}$ state or the $\ket{\text{vac}}$ 
state.

In order to discriminate between the states $\ket{+}$ and $\ket{\text{vac}}$, we measure the photon a second time.  Before this 
second measurement a phase rotation is applied to the photon, taking $\ket{\pm} \leftrightarrow \ket{\mp}$.  Therefore, if the 
state entering the measurement network is the vacuum, {\em both} modules will be measured in the $\ket{+}_a$ state. 
Assuming that measurements of the photonic module are error free, we are therefore able to uniquely discriminate the actual 
photonic states, $\ket{\pm}$ {\em and} the loss channel $\ket{\text{vac}}$.

Although the above scheme allows us to uniquely identify loss in the network, we also need to check that it is still effective 
when each of the two photonic modules are subjected to measurement errors and when loss can occur between the 
two measurements.  
We can summarize the possible measurement outcomes for the two modules under these error channels.
\begin{table}[ht!]
\begin{center}
\vspace*{4pt}   
\begin{tabular}{c|c}
Scenario & Results \\
\hline
$\ket{\text{vac}}$ & $\ket{+}_{M_1}\ket{+}_{M_2}$ \\
$\ket{+}$  \& error on $M_2$ & $\ket{+}_{M_1}\ket{+}_{M_2}$\\
$\ket{-}$  \& error on $M_1$ &  $\ket{+}_{M_1}\ket{+}_{M_2}$\\
$\ket{+}$  \& loss after $M_1$ & $\ket{+}_{M_1}\ket{+}_{M_2}$\\
$\ket{-}$  \& error on $M_1$ \& loss after $M_1$ & $\ket{+}_{M_1}\ket{+}_{M_2}$\\
\hline
$\ket{+}$ & $\ket{+}_{M_1}\ket{-}_{M_2}$\\
$\ket{-}$  \& error on $M_1$ \& $M_2$ & $\ket{+}_{M_1}\ket{-}_{M_2}$\\
$\ket{\text{vac}}$  \& error on $M_2$ &  $\ket{+}_{M_1}\ket{-}_{M_2}$\\
$\ket{+}$  \& error on $M_2$ \& loss after $M_1$ &  $\ket{+}_{M_1}\ket{-}_{M_2}$\\
$\ket{-}$  \& errors on $M_1$ \& $M_2$ \& loss after $M_1$ &  $\ket{+}_{M_1}\ket{-}_{M_2}$\\
\hline
$\ket{-}$  & $\ket{-}_{M_1}\ket{+}_{M_2}$\\
$\ket{+}$  \& error on $M_1$ \& $M_2$ & $\ket{-}_{M_1}\ket{+}_{M_2}$\\
$\ket{\text{vac}}$  \& error on $M_1$ &  $\ket{-}_{M_1}\ket{+}_{M_2}$\\
$\ket{+}$  \& error on $M_1$ \& loss after $M_1$ &  $\ket{-}_{M_1}\ket{+}_{M_2}$\\
$\ket{-}$  \& loss after $M_1$ &  $\ket{-}_{M_1}\ket{+}_{M_2}$\\
\hline
$\ket{-}$  \& error on $M_2$& $\ket{-}_{M_1}\ket{-}_{M_2}$\\
$\ket{+}$  \& error on $M_1$ & $\ket{-}_{M_1}\ket{-}_{M_2}$\\
$\ket{\text{vac}}$  + errors on $M_1$ and $M_2$ &  $\ket{-}_{M_1}\ket{-}_{M_2}$\\
$\ket{+}$  \& errors on $M_1$ \& $M_2$ \& loss after $M_1$ &  $\ket{-}_{M_1}\ket{-}_{M_2}$\\
$\ket{-}$  \& error on $M_2$ \& loss after $M_1$ &  $\ket{-}_{M_1}\ket{-}_{M_2}$\\
\end{tabular}
\caption{Possible physical scenarios corresponding to the measurements seen on each of the 
two photonic modules, $M_1$ and $M_2$.  For three of the four measurement patterns, the correct 
state is listed first.} 
\label{tab:errors}
\end{center}
\end{table} 

For each case, except $\ket{-}_{M_1}\ket{-}_{M_2}$, the correct 
error free result is listed first.  Ensuring correct fault-tolerant operation of the perpetual network requires protecting 
the system from the following,
\begin{enumerate}
\item Lost photons need to be reliably detected and replaced.  Additional errors may cause temporally correlated 
loss events, but these should not persist in the network (unless additional errors occur).
\item Two photons should {\em never} be injected into the preparation 
network at the same time.  Hence we tolerate a small increase in the correlated loss rate in order 
to completely suppress this possibility.
\end{enumerate}
Therefore, we always assume that no loss event has taken place if the modules are measured in the states 
$\ket{+}_{M_1}\ket{-}_{M_2}$ or $\ket{-}_{M_1}\ket{+}_{M_2}$ and the photon is rerouted back to the start of the preparation network. 
The other possibilities, if additional errors have occurred, are,
\begin{enumerate}
\item The wrong state was measured, but the photon is still present in the network.  This is a standard measurement error which is 
effectively corrected by the properties of the cluster.
\item A $\ket{\text{vac}}$ state was re-routed back into the network.  This causes a temporally correlated loss event but is corrected (with 
high probability) in the next cycle.
\item A $\ket{\text{vac}}$ state is re-routed back into the network combined with a measurement error.  The topological cluster corrects 
the measurement error and a correlated loss error is corrected (with high probability) in the next cycle.
\end{enumerate}
If the modules in the detection network are measured in the state $\ket{+}_{M_1}\ket{+}_{M_2}$ or $\ket{-}_{M_1}\ket{-}_{M_2}$ then 
we {\em do not} re-route back into the network and instead re-inject a new photon from the source (if one is available).  The 
adverse effect, when other errors occur, is that a photon may be unintentionally 
removed from the computer.  Therefore, some measurement errors in the detection network generate two correlated errors; the 
initial measurement error and assuming a photon is not re-injected from the source, the original photon 
is accidentally removed.  

These rules for interpreting the measurement results from the detection network ensures that lost photons can be 
detected and replaced, single failure events (in the detection modules) propagate to {\em at 
most} two temporally correlated errors and that two photons are never injected into the preparation network at the same time.

\subsection{Photon re-routing.}

\begin{figure*}[ht!]
\begin{center}
\resizebox{175mm}{!}{\includegraphics{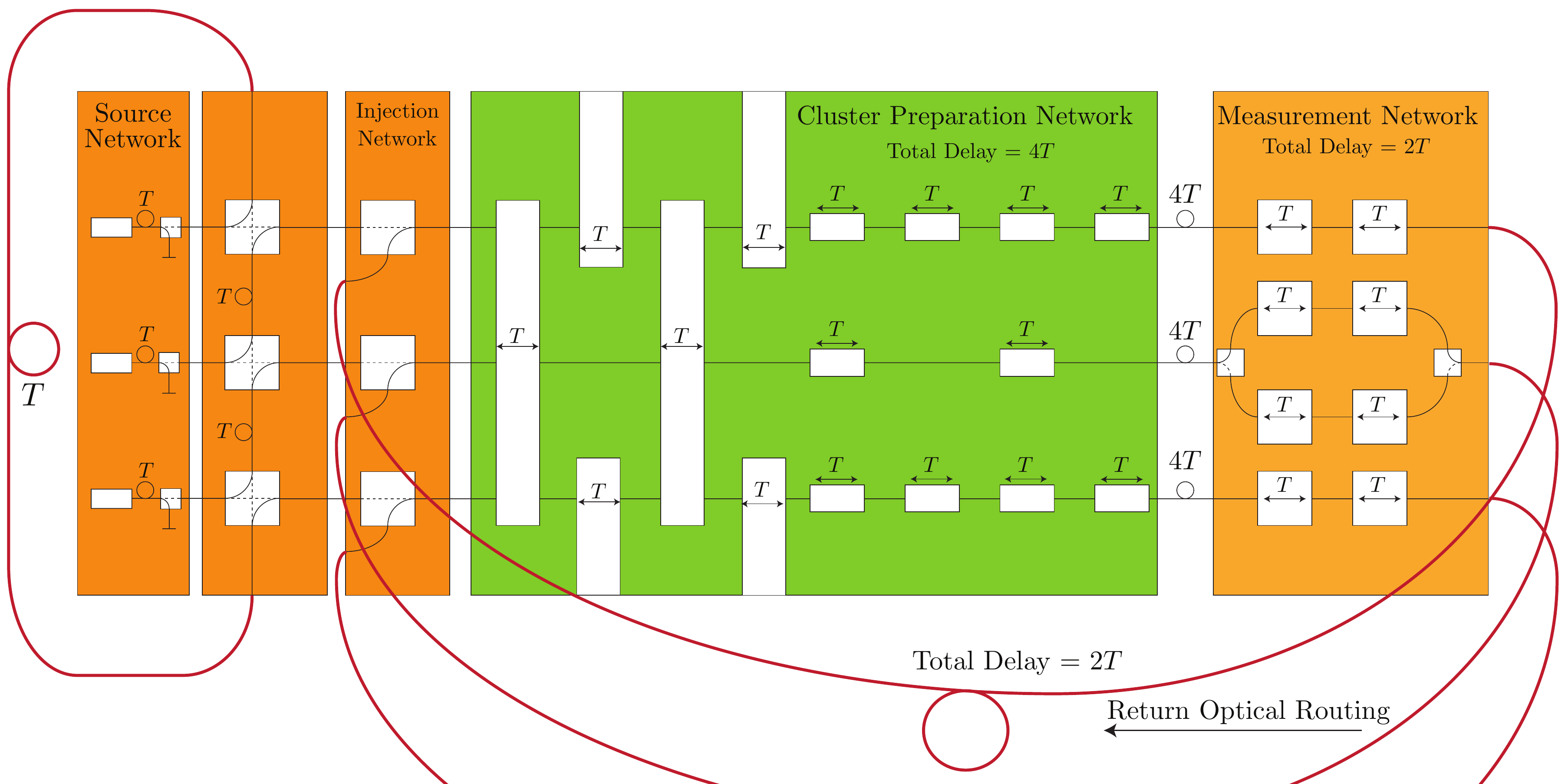}}
\end{center}
\vspace*{-10pt}
\caption{Cross sectional structure of the perpetual network.  The network contains three separate sub-networks.  The source network 
consists of an array of photonic modules which probabilistically distill a weak coherent pulse into single photons.  The injection 
network accepts photons from the source or photons recycled from the measurement network.  The cluster preparation network 
deterministically couples photons and was detailed previously in Ref.~\cite{DFSG08}.  Finally the measurement network which 
measures photons for computation, detects loss events and reroutes photons successfully measured back to the injection network.}
\label{fig:network3}
\end{figure*}

Once measured by the detection network, photons are re-routed back into the cluster preparation network.  The full structure of the 
perpetual network is illustrated in Fig.~\ref{fig:network3}.  The injection and source networks will be discussed shortly. 
The return optical routing of photons contains a total delay of $2T$.  This delay is required to ensure 
that re-injected photons maintain the temporal arrangement of the network.  Each photon spends a total of $10T$ within the cluster preparation 
network and the detection network.  Once a given photon is measured by the second module in the detection system, it must be further 
delayed by $2T$.  This is due to the fact that the atomic system needs to be measured, taking time $T$, and to allow 
the {\em classical} measurement signal to be transmitted from the detector network to the source network (once again taking time $T$).
The photon repetition rates for the network are $4T$ and $2T$ respectively.  Therefore, to maintain the temporal arrangement within 
the network, the total cycle time, $T_{T}$, must satisfy $T_{T}\mod (4T)= T_{T}\mod (2T) = 0$.  For the network in Fig.~\ref{fig:network3} 
the total cycle time is $T_{T} = 12T$, satisfying these conditions.  Consequently, photons arriving from the return path will re-enter the network 
at the correct temporal location.  
 
\section{Source and Injection Network}
\label{sec:injection}
The final part of the perpetual design is the injection network.  This part of the system accepts photons being re-routed back from the 
measurement network and also accepts photons from the network of probabilistic sources.  
As explained in Sec.~\ref{sec:sources}, each source distills single photons at random, but heralded, times and the rest of the computer 
will lose photons at random, but heralded, times.  Therefore, the injection network needs to be designed in a way that photons prepared 
by the sources can be routed to various injection points in the preparation network as they are required.  

This can be done by connecting each source to the optical lines immediately above.  Probabilistic sources are 
therefore linked together to form a uni-directional, linear nearest 
neighbor network (uLNN).  This network connecting sources to the preparation network is referred to as the {\em shunting} network.  
Individual wafers of photonic modules (illustrated in Fig.~\ref{fig:computer}b) have independent shunting networks and are 
{\em not} interconnected between separate wafters.  Each of the shunting networks have a size along the length of each wafer 
related to the fundamental probability that an individual source successfully prepares 
a single photon.  The size of the shunting network is given by $N = O(1/p_s)$, where $N$ is the number of sources connected in the 
network and $p_s$ is the 
probability of distilling a single photon, per cycle.  If photonic modules are utilized as probabilistic sources, then the cycle time is $3T$: $2T$ 
for module initialization and measurement and an interaction time of $T$.   
The size of the source network is chosen such that, on average, one photon is successfully distilled within the network every $3T$.
The source network is connected to the shunting network via a delay of $T$, this gives sufficient time for each module to be measured 
to confirm if a single photon has been successfully distilled before being introduced into the injection network.  

The shunting network connects each optical line to the one above it with an additional time delay of $T$.  The boundary conditions are 
periodic, with the uppermost optical line connected to the lowermost optical line.  Given that $T$ is defined via the operational time 
of a photonic module (between 10ns and 1$\mu$s), the spatial extent of the shunting network is 
limited to between 1-100m, more than sufficient for a large scale array.
\begin{figure}[ht!]
\begin{center}
\resizebox{50mm}{!}{\includegraphics{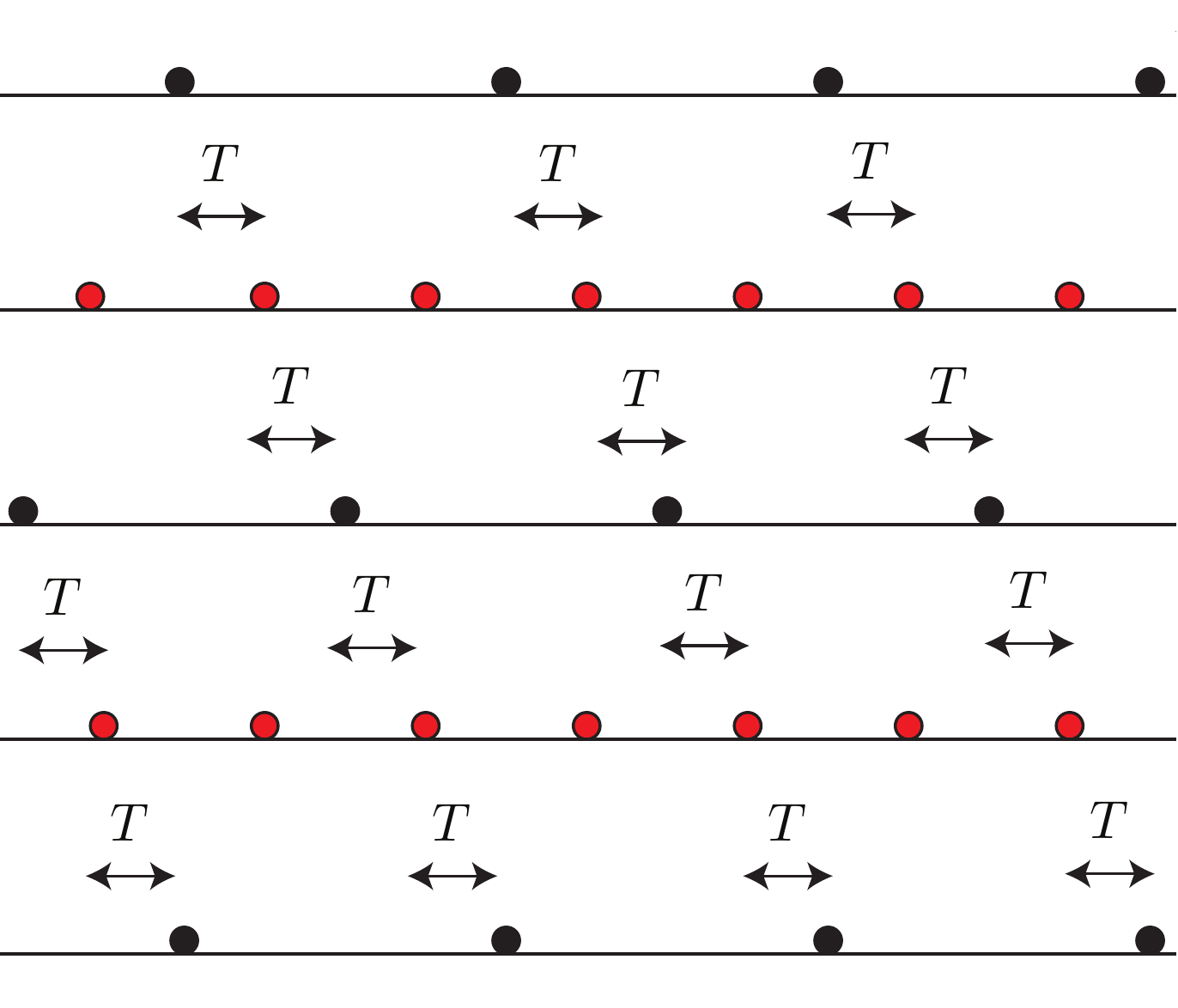}}
\end{center}
\vspace*{-10pt}
\caption{Temporal arrangement of photons entering the cluster preparation network.  Photons in neighboring optical lines 
are separated by time $T$.  Hence if a photon is shunted one optical line upwards, it must also be delayed by $T$ to be placed 
in the correct location.}
\label{fig:temp}
\end{figure}
The required time delay of $T$ between neighboring optical lines is due to the temporal arrangement of photons entering the preparation 
network, Fig.~\ref{fig:temp} illustrates.  Each individual optical line has a pulse separation of either $2T$ or $4T$ and each pulse in neighboring 
optical lines is separated by $T$.  If a source produces a heralded photon at the appropriate time to be injected to its corresponding 
optical line, but the photon is not required, it is then routed one line up.  The delay of $T$ ensures this shunted 
photon is at the right temporal location should the photon be required to replace a loss event in the next optical line.  

\begin{figure*}[ht!]
\begin{center}
\resizebox{165mm}{!}{\includegraphics{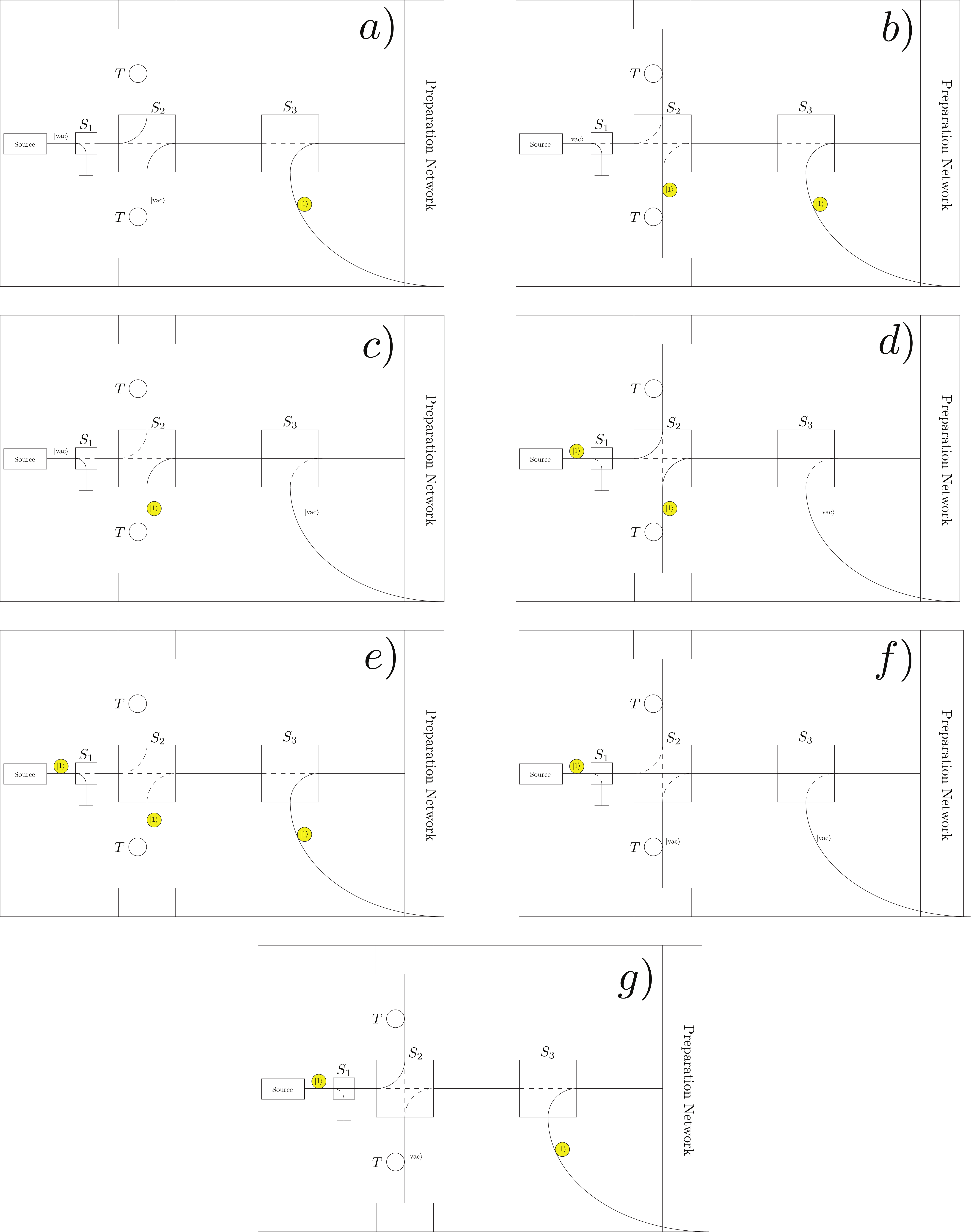}}
\end{center}
\vspace*{-10pt}
\caption{Each of the seven switching scenarios dictating how photons are either recycled from the measurement network, accepted in 
or out of the shunting network or accepted from the source.  In these figures the $\ket{\text{vac}}$ states have {\em always} been detected 
by the heralding associated with the sources or the detectors.  Vacuum states can also be present which have not been detected.  For 
example, a photon is lost while it is being rerouted from the detectors back to the injection network.  In this case, the switching 
pattern assumes the photon is still present, re-injecting a $\ket{\text{vac}}$ state into the computer which will be 
corrected in the next cycle.}
\label{fig:switching}
\end{figure*}

The shunting network is designed to cycle freshly prepared photons in a loop until they are required.  Each time a loss event is confirmed by the measurement network, the system accepts a photon from the shunting network, if one is available.  For each optical line, there are seven 
distinct switching scenarios which are illustrated in Fig.~\ref{fig:switching}.
These switching scenarios represent all cases, where photons are successfully recycled, where the source successfully 
distill a state and where a photon is present in the shunting network.  Each pattern can be briefly summarized,
\begin{enumerate}
\item {\bf Subfigure a)}: Photon is successfully recycled and simply rerouted back into the preparation network.
\item {\bf Subfigure b)}: Photon is successfully recycled and a second photon is present in the shunting network.  Recycled photon 
is routed back into the preparation network and the photon in the shunting network is routed up into the next optical line.
\item {\bf Subfigure c)}: Photon is confirmed as lost and a second photon is present in the shunting network.  The photon in 
the shunting network is routed into the preparation network.
\item {\bf Subfigure d)}: Photon is confirmed as lost at the same time as a second photon is present in the shunting network and 
a new photon is successfully distilled from the source.  The photon in the shunting network is routed into the preparation network 
and the photon from the source enters the shunting network.
\item {\bf Subfigure e)}: All three photons are present.  The photon from the source is routed to a termination point and removed from the 
network, the other two photons are routed in the same way as subfigure b).
\item {\bf Subfigure f)}: Photon is confirmed as lost and source successfully distills a photon.  This new photon is direcly routed to the 
preparation network.
\item {\bf Subfigure g)}: Photon is successfully recycled and the source distills a photon.  The recycled photon is rerouted back into the 
preparation network and the newly distilled photon is routed into the shunting network.
\end{enumerate}

Photon loss can occur at any point in this network and we do not assume any non-demolition ``probing" of the network to confirm if 
photons are still present.  The only classical signals available are the heralding signal from each source and the signal from 
the measurement network confirming loss events.  As each distilled photon is injected into the shunting network, they are classically 
tracked.  If these photons are lost before being injected into the preparation network, then this will produce another loss event within 
the computer which, with high probability, will be corrected in the next cycle.  
 
\section{Network Simulations}
\label{sec:simulations}
To confirm that this network design operates as intended, direct numerical simulations were performed.  It should be stressed that we are 
{\em not} simulating any quantum aspect of this architecture.  We are simply focussing on the network structure under finite photon 
loss and low probability source injection.  

As detailed in previous sections, there are a total of $12T$ time steps within the network, $4T$ in the cluster preparation 
network, $2T$ in the measurement network, a $4T$ delay between the preparation and measurement networks and a $2T$ delay 
when rerouting photons.  During each of these steps, individual 
photons are subjected to loss with a probability of $p_L$.  We refer to $p_L$ as the {\em per component loss probability}.  We 
also define $p_c$ as the {\em per cycle loss probability}, this is the probability that a photon is lost between entering the 
preparation network and re-entering it again after it is recycled, $p_c = 1-(1-p_L)^{12} \approx 12p_L$.  

Within the source network, each photon is delayed by $T$ immediately after it is distilled (in order for the module to be measured, confirming 
distillation) and again subjected to loss with probability $p_L$.  Once a photon enters the shunting network it is continually shunted in a 
loop, each step in the loop takes time $T$.  The shunting continues until the photon is accepted into the network or lost.  

In each simulation we vary the total time the network is running and the bias between $p_L$ and the probability of success for 
each source, $p_s$.  The source probability is fixed and dictated by the size of the network.  For a network 
consisting of $N$ optical lines, the source probability is $p_s = 1/(3N)$, guaranteeing that on average one photon is successfully 
distilled every $3T$ steps.  The simulations are designed to determine the following,
\begin{enumerate}
\item For a given network size $N=1/(3p_s)$ and bias $B=p_s/p_L > 1$, does the network saturate?  
i.e. how large is $B$ such that as $t \rightarrow \infty$ each temporal position in the network occupied with a single photon. 
\item For a given network size $N=1/(3p_s)$ and a bias $B = p_s/p_L$, how many timesteps are required before the network is saturated 
with photons.  Essentially how long is required to boot-up the computer before computation can proceed.  
\end{enumerate}

For $N$ optical lines, the total 
number of photons required to saturate the network is given by $P = 9N/2$.  For a large scale computer, consisting of $N\times N$ optical 
lines, the total number of photons required to boot-up the computer is $P=81N^2/4$. 
However, as each of the individual uLNN shunting networks in the computer are {\em not} interconnected, it is sufficient to simulate 
a single cross-section containing $N$ optical lines.  

\section{Results}
\label{sec:results}
The numerical simulations are performed varying the network size $N$, the bias $B$ and the total number of timesteps $t$.  The source 
and loss probabilities are given by $p_s = 1/(3N)$ and $p_L = p_s/B = 1/(3BN)$.  
Monte-Carlo simulations were performed using $10^3$ statistical runs.  
Due to available computation power, we have simulated up to 168 optical lines.  This corresponds to a source 
probability as low as $p_s = 2\times 10^{-3}$ and hence an approximate error on the distilled single photon state of $\epsilon = 1-F = 
6.7\times 10^{-7}$.  Shown in Fig.~\ref{fig:simtimes} is the plot for $N=88$, with other simulation results shown in Appendix.~\ref{sec:AppA}.  
This simulation illustrates the total number of photons present in the network as a function 
of $t$ for $ 2 \leq B \leq 192$ averaged over $10^3$ statistical runs.  

The first thing to notice is that the total number of photons in the network is greater than $P = 9N/2$ when $B$ and $t$ are large.  This is 
due to the shunting network.  While the {\em computer} requires $9N/2$ photons to saturate, additional photons are also present in the 
shunting network.  These photons are used to replace confirmed loss events while the computer is in operation.  On average, when 
saturated, the shunting network will contain an extra $N/2$ photons.  These two bounds are illustrated on each plot.  As $B\rightarrow \infty$, 
the total number of photons in the network approaches $P = 5N$.  
\begin{figure}[ht!]
\begin{center}
\resizebox{85mm}{!}{\includegraphics{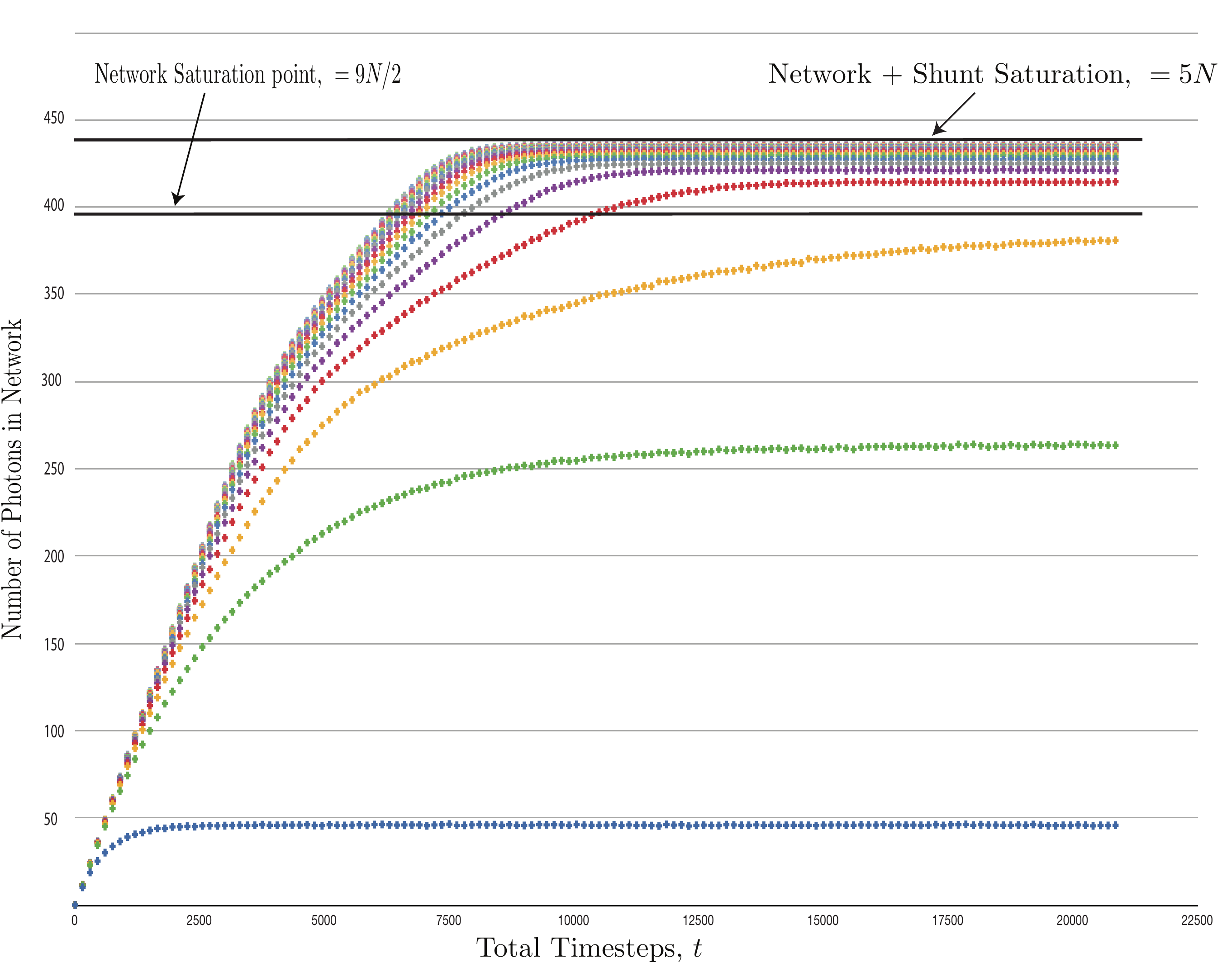}}
\end{center}
\vspace*{-10pt}
\caption{Simulations showing the total number of photons in the network as a function of total timesteps for 
$N = 88$ optical lines.  Each curve 
represents higher values of $2\leq B\leq 192$ in increments of ten.  
Two bounds exists in the system.  The computational network requires $9N/2$ photons 
to saturate the system.  The second bound represents additional photons present in the shunting network.  In total $5N$ photons 
are required to saturate the entire system.  Other results for $8 \leq N \leq 168$ are shown in Appendix.~\ref{sec:AppA}}
\label{fig:simtimes}
\end{figure}

The second set of simulations shown in Appendix.~\ref{sec:AppB} (with Fig.~\ref{fig:simbias} illustrating for $N=88$) 
examines the total number of photons present in the network as a function of 
$B$.  Each data point is taken at the maximum value of $t$ simulated in Appendix.~\ref{sec:AppA}.   
For each value of $N$, the network 
saturates once the bias reaches $B\approx 30$.  Note that this {\em threshold} bias does decrease slightly as $N$ increases, 
but it is essentially independent of the network size.  A bias threshold 
of 30 translates to a {\em per cycle} loss probability of $p_c = 0.4p_s$.  Hence, provided that the source probability is 
approximately 2.5 times higher than the {\em per cycle} loss probability of each photon, 
the system will eventually saturate and computation can proceed.
\begin{figure}[ht!]
\begin{center}
\resizebox{85mm}{!}{\includegraphics{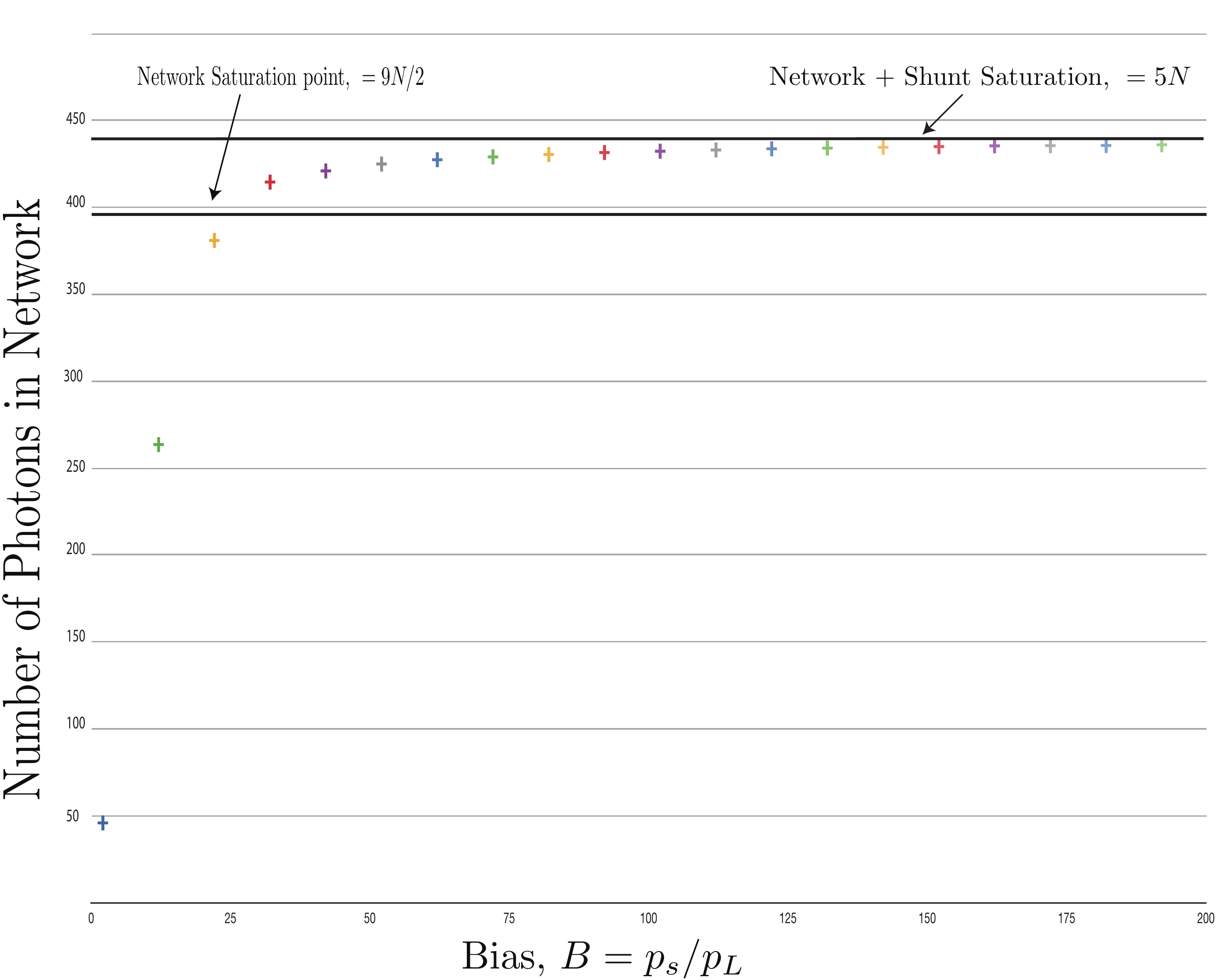}}
\end{center}
\vspace*{-10pt}
\caption{Simulations showing the total number of photons in the network as a function of $B$ for 
$N = 88$ optical lines. Each data point is taken at the maximum value of $t$ simulated in Fig.~\ref{fig:simtimes}.  
The system saturates at $B\approx 30$.  Other results for $8 \leq N \leq 168$ are shown in Appendix.~\ref{sec:AppB}, 
illustrating that this bias threshold is essentially constant for all $N$.}
\label{fig:simbias}
\end{figure}

\begin{figure}[ht!]
\begin{center}
\resizebox{85mm}{!}{\includegraphics{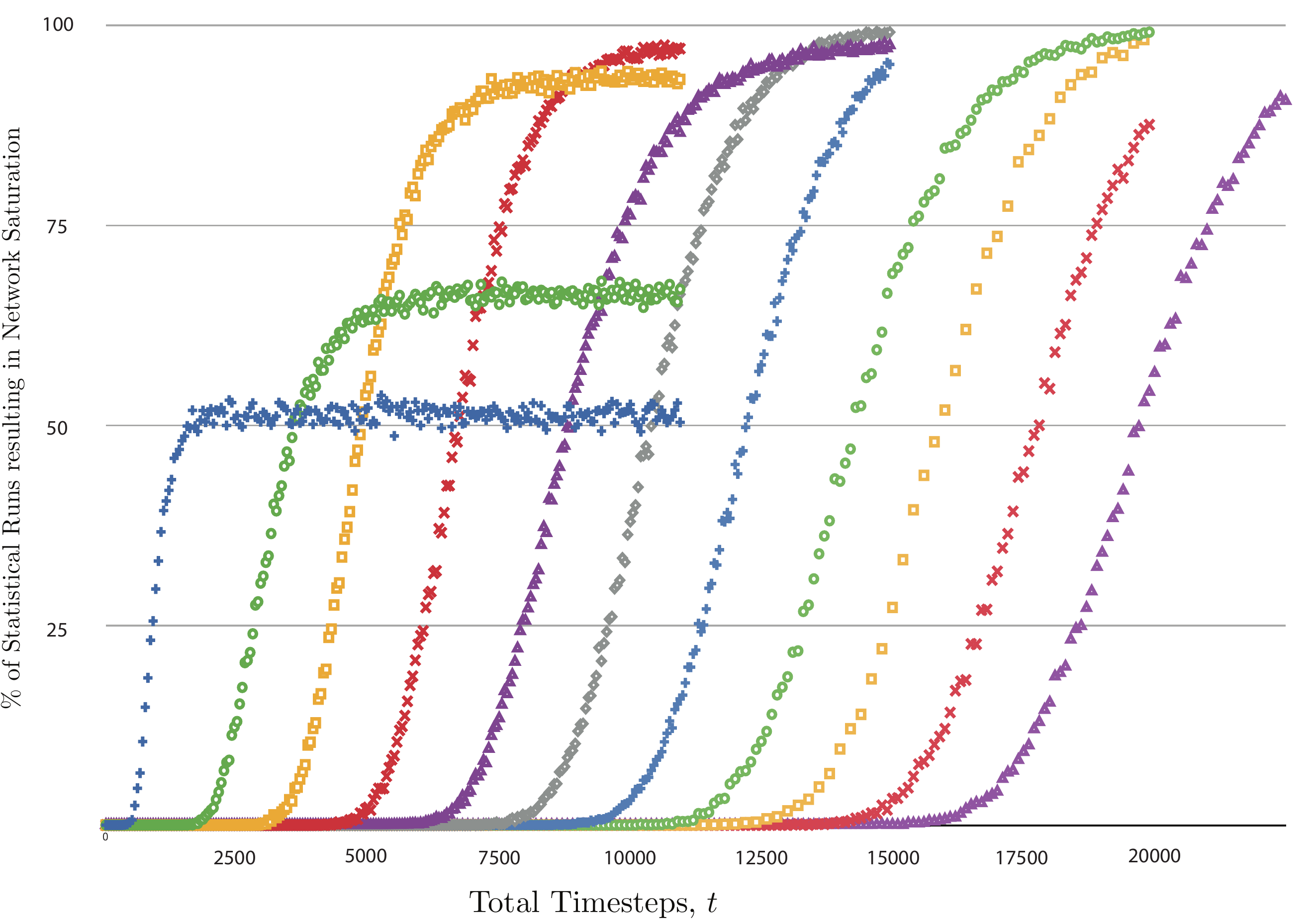}}
\end{center}
\vspace*{-10pt}
\caption{Percentage of statistical samples ($3\times 10^3$ total) 
which result in network saturation as a function of total timesteps, $t$, and network size, $N$.  
For each simulation we fix the bias at $B=32$.  Each curve shows a clear transition when the computational network saturates 
and can be used for computation.  The point at which approximately 50\% of statistical samples result in $>9N/2$ photons 
being present in the network is the boot-up time illustrated in Fig.~\ref{fig:bootup}.  At $B=32$, we are sitting very close to the 
bias threshold for low values of $N$.  This is why the curves, especially for $N=8$ and $N=24$,  
show less than 100\% of the samples saturating for large $t$.}
\label{fig:bootup2}
\end{figure}

Finally, we can determine an approximate boot-up time for the computer.  For this, we fix the bias at $B=32$ and 
re-simulate the network for a total of $3\times 10^3$ samples.  We then plot as a function of total timesteps, the percentage of 
all samples which result in complete saturation of the preparation network. Fig.~\ref{fig:bootup2}, illustrates.

\begin{figure}[ht!]
\begin{center}
\resizebox{85mm}{!}{\includegraphics{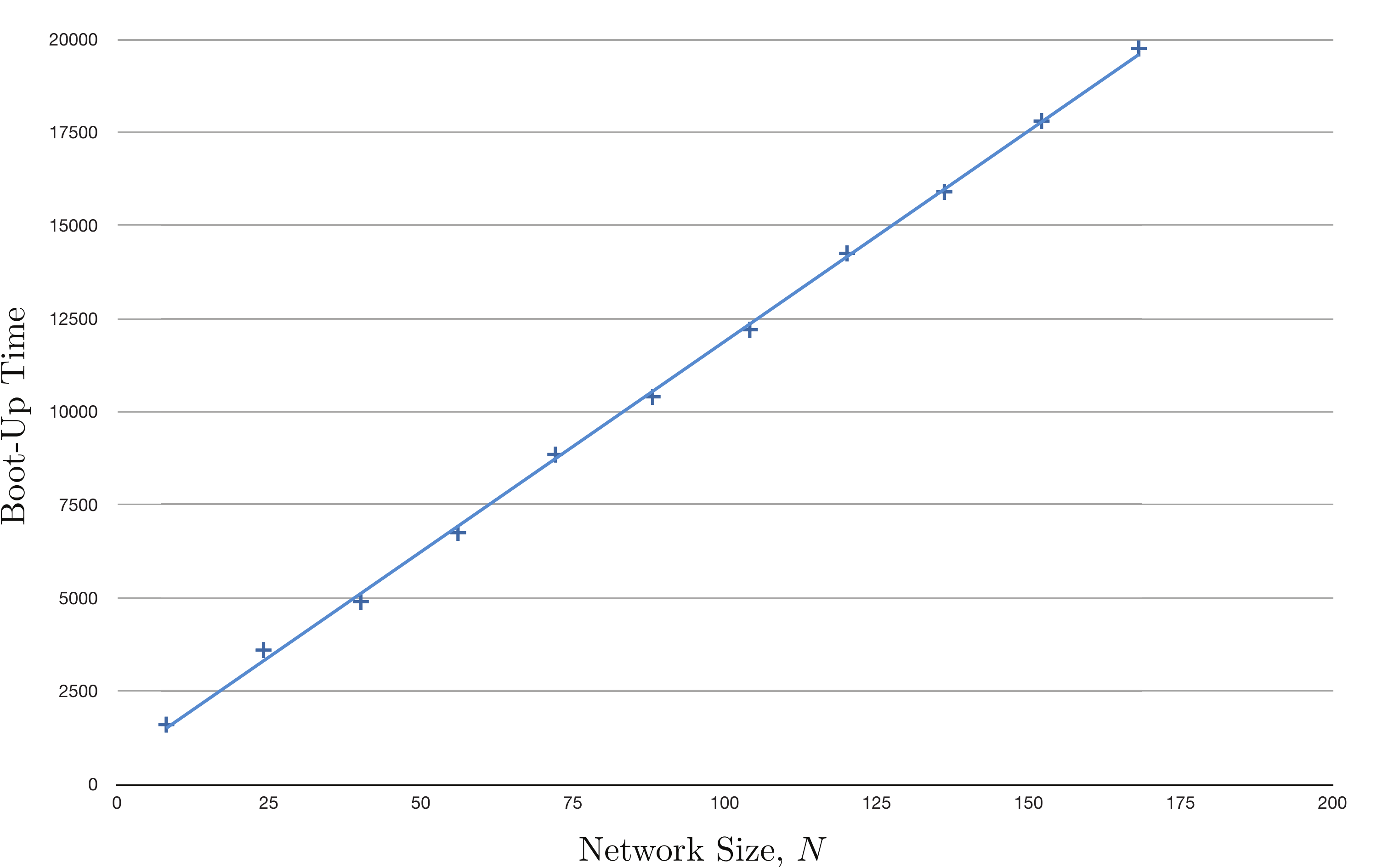}}
\end{center}
\vspace*{-10pt}
\caption{Scaling of boot-up time as a function of network size, $N$.  Each point was estimated from the simulations in Appendix.~\ref{sec:AppA} 
at a bias of $B=32$, finding the minimum number of timesteps such that 50\% of the simulations resulted in $> 9N/2$ photons being present in
 the network.  The scaling of boot-up time scales linearly with $N$, hence inversely with $p_s$.  A more conservative estimate 
 for the boot-up time for each datapoint results in a small constant shift in the above curve which, for large $N$, does not significantly change 
 our estimates.}
\label{fig:bootup}
\end{figure}

Each curve shows a clear transition, where the total timesteps is large enough such that the network saturates~\footnote{As we set $B=32$, 
we are sitting very close to the bias threshold for small values of $N$, hence for large $t$ not all samples result in network 
saturation.  This is why the curves, especially for $N=8$ and $N=24$,   show less than 100\% saturation for large $t$}.
For estimating the boot-up time for various network sizes, we simply find the approximate point where 50\% of all samples lead 
to $> 9N/2$ photons in the network.  
Fig.~\ref{fig:bootup} illustrates the approximate boot-up time for $8\leq N\leq 168$.

As you can see, there is a clear linear relationship between the boot-up time and the total size of the network.   As the size of the 
network is dictated by the source probability, $N = 1/(3p_s)$, the time required to saturate the network scales inversely with $p_s$.  
While, even for modestly sized networks, the total number of timesteps required for boot-up is large, the actual physical time is more than 
acceptable.  Each timestep, $T$, represents the 
operational time for a photonic module.  Hence assuming $T = 1\mu s$ the scaling, in units of milliseconds, 
is approximately,
\begin{equation}
t_{\text{boot-up}} (\text{ms}) \approx \frac{1}{30p_s}.
\end{equation}

We can now summarize the requirements for a perpetual design running with highly probabilistic sources,
\begin{table}[ht!]
\begin{center}
\vspace*{4pt}   
\begin{tabular}{c|c|c|c}
$p_s$ & Max-$p_c$ & $t_{\text{bootup}}$ ($\approx$) & shunt network size, $N$, ($\approx$)\\
\hline
$10^{-2}$ & $4\times 10^{-3}$ & 3 ms & 30\\
$10^{-3}$ & $4\times 10^{-4}$ & 30ms & $3\times 10^{2}$\\
$10^{-4}$ & $4\times 10^{-5}$ & 300ms & $3\times 10^{3}$\\
$10^{-5}$ & $4\times 10^{-6}$ & 3s & $3\times 10^{4}$\\
$10^{-6}$ & $4\times 10^{-7}$ & 30s & $3\times 10^{5}$
\end{tabular}
\caption{Estimation of boot-up times and shunting network sizes for various values of distillation probability, $p_s$.  Max-$p_c$ is 
the maximum {\em per cycle} probability of photon loss such that the computational network will saturate (corresponding to a Bias 
of approximately $p_s = 2.5p_c$.} 
\label{tab:errors}
\end{center}
\end{table} 

A reasonable parameter range for topological cluster states requires a per cycle loss probability, $p_c$, 
between approximately $10^{-2}$ and $10^{-4}$.  
Higher loss probabilities begin to require unreasonable cluster resources to correct~\cite{BS10} while loss probabilities lower than 
$\approx 10^{-4}$ are experimentally unrealistic.  Therefore, source probabilities can be in the range of approximately 
$2.5\% \rightarrow 0.025\%$ with each shunting network connecting 10 $\rightarrow 300$ optical lines and an 
boot-up time of between 1 $\rightarrow$ 100ms.

These estimates represent approximate upper bounds for the boot-up time and the size of each shunting network.  
If each probabilistic source succeeds with a much higher probability (while loss 
probabilities in the network remain fixed), then both the boot-up time and the size of each shunting network will decrease.

\section{Conclusions}
This discussion has shown how highly probabilistic photon sources can be integrated into a scalable optical architecture.  
By replacing deterministic sources and detectors with the photonic module, this modified design is constructed exclusively from a 
single quantum component.  These results equally apply to any other form of highly probabilistic photonic sources, provided they can 
be integrated into a computational network similar to the optical architecture used in this work.  

The introduction of the shunting network, connecting probabilistic sources to the cluster preparation network, combined 
with the ability to recycle measured photons essentially leads to 
a pseudo set of on demand sources.  It is easily shown that by having sources with a slightly higher probability of 
success than loss, the system will be able to compensate photon loss during computation.  The shunting network solved 
the problem of being able to route photons prepared in random locations in the network to other random locations where photons 
are lost.  This structure effectively achieved the same goal as extremely complicated multi-port switches and extensive optical delay. 

This analysis demonstrates the practicality of a limited amount of highly probabilistic quantum components in a large scale architecture.  
Unlike other results, illustrating the theoretical ability to construct a quantum computer from probabilistic components, this work 
has maintained the explicit architectural structure of the optical computer.  By starting from a fully deterministic architecture, we could 
carefully integrate a limited amount of probabilistic technology without sacrificing the overall structure of the system.  
The next major step is to determine if additional probabilistic technology can replace other currently deterministic components.  The 
results presented in this paper give us a certain amount of optimism that this can be achieved.

\section{Acknowledgements}
The authors wish to 
acknowledge the support of MEXT, JST and FIRST projects.

\bibliographystyle{unsrt}
\bibliography{bib1}  

\begin{thebibliography}{10}

\bibitem{CZ95}
J.I. Cirac and P.~Zoller.
\newblock {Quantum Computations with Cold Trapped Ions}.
\newblock {\em Phys. Rev. Lett.}, 74:4091, 1995.

\bibitem{CFH97}
D.G. Cory, A.F. Fahmy, and T.F. Havel.
\newblock {Ensemble Quantum Computing by NMR Spectroscopy}.
\newblock {\em Proc. National Academy of Science}, 94:1634--1639, 1997.

\bibitem{GC97}
N.~Gershenfeld and I.L. Chuang.
\newblock {Bulk Spin Resonance Quantum Computing}.
\newblock {\em Science}, 275:350, 1997.

\bibitem{K98}
B.E. Kane.
\newblock {A Silicon-Based nuclear spin Quantum Computer}.
\newblock {\em Nature (London)}, 393:133, 1998.

\bibitem{LD98}
D.~Loss and D.P. DiVincenzo.
\newblock {Quantum Computation with quantum dots}.
\newblock {\em Phys. Rev. A.}, 57:120, 1998.

\bibitem{KLM01}
E.~Knill, R.~Laflamme, and G.J. Milburn.
\newblock {A Scheme for Efficient Quantum Computation with linear optics}.
\newblock {\em Nature (London)}, 409:46, 2001.

\bibitem{MOLTW99}
J.E. Mooij, T.P. Orlando, L.~Levitov, L.~Tian, C.H. van~der Wal, and S.~Lloyd.
\newblock {Josephson Persistent-Current Qubit}.
\newblock {\em Science}, 285:1096--1039, 1999.

\bibitem{NPT99}
Y.~Nakamura, Yu.~A. Pashkin, and J.S. Tsai.
\newblock {Coherent control of Macroscopic Quantum states in a Cooper-Pair
  box}.
\newblock {\em Nature (London)}, 398:786, 1999.

\bibitem{ARPA}
ARDA.
\newblock {Quantum information science and technology roadmap project,
  http://qist.lanl.gov}, 2004.

\bibitem{KMNRDM07}
P.~Kok, W.J. Munro, K.~Nemoto, T.C. Ralph, J.P. Dowling, and G.J. Milburn.
\newblock {Linear Optical quantum computing}.
\newblock {\em Rev. Mod. Phys.}, 79:135, 2007.

\bibitem{CNHM03}
I.~Chiorescu, Y.~Nakamura, C.J.P.M Harmans, and J.E. Mooij.
\newblock {Coherent quantum dynamics of a superconducting flux qubit}.
\newblock {\em Science}, 299:1869, 2003.

\bibitem{YPANT03}
T.~Yamamoto, Yu.~A. Pashkin, O.~Astafiev, Y.~Nakamura, and J.S. Tsai.
\newblock {Demonstration of a conditional gate operation using superconducting
  charge qubits}.
\newblock {\em Nature (London)}, 425:941, 2003.

\bibitem{CLW04}
J.~Chiaverini, D.~Leibfried, T.~Schaetz, M.D. Barrett, R.B. Blakestad,
  J.~Britton, W.M. Itano, J.D. Jost, E.~Knill, C.~Langer, R.~Ozeri, and D.J.
  Wineland.
\newblock {Realization of quantum error correction}.
\newblock {\em Nature (London)}, 432:602, 2004.

\bibitem{HHB05}
H.~H\"{a}ffner, W.~H\"{a}nsel, C.F. Roos, J.~Benhelm, D.~Chek al~kar,
  M.~Chwalla, T.~K\"{o}rber, U.D. Rapol, M.~Riebe, P.O. Schmidt, C.~Becher,
  O.~G\"{u}hne, W.~D\"{u}r, and R.~Blatt.
\newblock {Scalable multiparticle entanglement of trapped ions}.
\newblock {\em Nature (London)}, 438:643, 2005.

\bibitem{GHW05}
J.~Gorman, D.G. Hasko, and D.A. Williams.
\newblock {Charge-Qubit operation of an Isolated double quantum dot}.
\newblock {\em Phys. Rev. Lett.}, 95:090502, 2005.

\bibitem{G06}
T.~Gaebel, M.~Domhan, I.~Popa, C.~Wittmann, P.~Neumann, F.~Jelezko, J.R.
  Rabeau, N.~Stavrias, A.D. Greentree, S.~Prawer, J.~Meijer, J.~Twamley, P.R.
  Hemmer, and J.~Wrachtrup.
\newblock {Room Temperature coherent control of coupled single spins in solid}.
\newblock {\em Nature Physics (London)}, 2:408, 2006.

\bibitem{H06}
R.~Hanson, F.M. Mendoza, R.J. Epstein, and D.D. Awschalom.
\newblock {Polarization and Readout of Coupled Single Spins in Diamond}.
\newblock {\em Phys. Rev. Lett.}, 97:087601, 2006.

\bibitem{G07}
M.V.~Gurudev Dutt, L.~Childress, L.~Jiang, E.~Togan, J.~Maze, F.~Jelezko, A.S.
  Zibrov, P.R. Hemmer, and M.D. Lukin.
\newblock {Quantum Register Based on Individual Electronic and Nuclear Spin
  Qubits in Diamond}.
\newblock {\em Science}, 316:1312, 2007.

\bibitem{OPWRB03}
J.L. O'Brien, G.J. Pryde, A.G. White, T.C. Ralph, and D.~Branning.
\newblock {Demonstration of an all optical quantum controlled-NOT gate}.
\newblock {\em Nature (London)}, 426:264, 2003.

\bibitem{DFSG08}
S.J. Devitt, A.G. Fowler, A.M. Stephens, A.D. Greentree, L.C.L. Hollenberg,
  W.J. Munro, and K.~Nemoto.
\newblock {Architectural design for a topological cluster state quantum
  computer}.
\newblock {\em New. J. Phys.}, 11:083032, 2009.

\bibitem{HFJR10}
D.A. Herrera-Marti, A.G. Fowler, D.~Jennings, and T.~Rudolph.
\newblock {A Photonic Implementation for the Topological Cluster State
  Computer}.
\newblock {\em Phys. Rev. A.}, 82:032332, 2010.

\bibitem{MLFY10}
R.~Van Meter, T.D. Ladd, A.G. Fowler, and Y.~Yamamoto.
\newblock {Distributed Quantum Computation Architecture Using Semiconductor
  Nonophotonics}.
\newblock {\em Int. J. Quant. Inf.}, 8:295, 2010.

\bibitem{DFTMN10}
S.J. Devitt, A.G. Fowler, T.~Tilma, W.J. Munro, and K.~Nemoto.
\newblock {Classical Processing Requirements for a Topological Quantum
  Computing Systems}.
\newblock {\em Int. J. Quant. Inf.}, 8:1, 2010.

\bibitem{JMFMKLY10}
N.~Cody Jones, R.~Van Meter, A.G. Fowler, P.L. McMahon, J.~Kim, T.D. Ladd, and
  Y.~Yamamoto.
\newblock {A Layered Architecture for Quantum Computing Using Quantum Dots}.
\newblock {\em arxiv:1010.5022}, 2010.

\bibitem{DMN08}
S.J. Devitt, W.J. Munro, and K.~Nemoto.
\newblock {High Performance Quantum Computing}.
\newblock {\em arxiv:0810.2444}, 2008.

\bibitem{FWHLMH10}
A.G. Fowler, D.S. Wang, C.D. Hill, T.D. Ladd, R.~Van Meter, and L.C.L.
  Hollenberg.
\newblock {Surface code quantum communication}.
\newblock {\em Phys. Rev. Lett.}, 104:180503, 2010.

\bibitem{MHSDN10}
W.J. Munro, K.A. Harrison, A.M. Stephens, S.J. Devitt, and K.~Nemoto.
\newblock {From quantum multiplexing to high-performance quantum networking}.
\newblock {\em Nature Photonics}, 4:792, 2010.

\bibitem{RH07}
R.~Raussendorf and J.~Harrington.
\newblock {Fault-tolerant quantum computation with high threshold in two
  dimensions}.
\newblock {\em Phys. Rev. Lett.}, 98:190504, 2007.

\bibitem{RHG07}
R.~Raussendorf, J.~Harrington, and K.~Goyal.
\newblock {Topological fault-tolerance in cluster state quantum computation}.
\newblock {\em New J. Phys.}, 9:199, 2007.

\bibitem{FG08}
A.G. Fowler and K.~Goyal.
\newblock {Topological cluster state quantum computing}.
\newblock {\em Quant. Inf. Comp.}, 9:721, 2009.

\bibitem{DGOH07}
S.J. Devitt, A.D. Greentree, R.~Ionicioiu, J.L. O'Brien, W.J. Munro, and L.C.L.
  Hollenberg.
\newblock {The Photonic Module: an on-demand resource for photonic
  entanglement}.
\newblock {\em Phys. Rev. A.}, 76:052312, 2007.

\bibitem{SNMK03}
S.~Scheel, K.~Nemoto, W.J. Munro, and P.L. Knight.
\newblock {Measurement-induced Nonlinearity in Linear Optics}.
\newblock {\em Phys. Rev. A.}, 68:032310, 2003.

\bibitem{SMENK06}
S.~Scheel, W.J. Munro, J.~Eisert, K.~Nemoto, and P.~Kok.
\newblock {Feed-forward and its role in conditional linear optical quantum
  dynamics}.
\newblock {\em Phys. Rev. A.}, 73:034301, 2006.

\bibitem{YR03}
N.~Yoran and B.~Reznik.
\newblock {Deterministic Linear Optics Quantum Computation with Single Photon
  Qubits}.
\newblock {\em Phys. Rev. Lett.}, 91:037903, 2003.

\bibitem{N04}
M.A. Nielsen.
\newblock {Optical quantum computation using cluster states}.
\newblock {\em Phys. Rev. Lett.}, 93:040503, 2004.

\bibitem{HGMR04}
A.J.F. Hayes, A.~Gilchrist, C.R. Myers, and T.C. Ralph.
\newblock {Utilizing encoding in scalable linear optics quantum computing}.
\newblock {\em J. Opt. B: Quantum Semiclass. Opt.}, 6:533, 2004.

\bibitem{BR05+}
D.E. Browne and T.~Rudolph.
\newblock {Resource-Efficient Linear Optical Quantum Computation}.
\newblock {\em Phys. Rev. Lett.}, 95:010501, 2005.

\bibitem{ND05}
M.A. Nielsen and C.M. Dawson.
\newblock {Fault-tolerant quantum computation with cluster states}.
\newblock {\em Phys. Rev. A.}, 71:042323, 2005.

\bibitem{DR05}
L.-M. Duan and R.~Raussendorf.
\newblock {Efficient Quantum Computation with Probabilistic Quantum Gates}.
\newblock {\em Phys. Rev. Lett.}, 95:080503, 2005.

\bibitem{DHN06}
C.M. Dawson, H.L. Haselgrove, and M.A. Nielsen.
\newblock {Noise thresholds for optical cluster-state quantum computation}.
\newblock {\em Phys. Rev. A.}, 73:052306, 2006.

\bibitem{GKE06}
D.~Gross, K.~Kieling, and J.~Eisert.
\newblock {Potential and limits to cluster-state quantum computing using
  probabilistic gates}.
\newblock {\em Phys. Rev. A.}, 74:042343, 2006.

\bibitem{GHR07}
A.~Gilchrist, A.J.F. Hayes, and T.C. Ralph.
\newblock {Efficient parity-encoded optical quantum computing}.
\newblock {\em Phys. Rev. A.}, 75:052328, 2007.

\bibitem{KRE07}
K.~Kieling, T.~Rudolph, and J.~Eisert.
\newblock {Percolation, renormalization, and quantum computing with
  non-deterministic gates}.
\newblock {\em Phys. Rev. Lett.}, 99:130501, 2007.

\bibitem{KGE07}
K.~Kieling, D.~Gross, and J.~Eisert.
\newblock {Cluster state preparation using gates operating at arbitrary success
  probabilities}.
\newblock {\em New. J. Phys.}, 9:200, 2007.

\bibitem{RB01}
R.~Raussendorf and H.-J. Briegel.
\newblock {A One way Quantum Computer}.
\newblock {\em Phys. Rev. Lett.}, 86:5188, 2001.

\bibitem{GPWRZ04}
S~Gasparoni, J.-W. Pan, P.~Walther, T.~Rudolph, and A.~Zeilinger.
\newblock {Realization of a Photonic Controlled-NOT Gate Sufficient for Quantum
  Computation}.
\newblock {\em Phys. Rev. Lett.}, 93:020504, 2004.

\bibitem{O07}
J.L. O'Brien.
\newblock {Optical Quantum Computing}.
\newblock {\em Science}, 318:1567, 2007.

\bibitem{LBYP07}
C.-Y. Lu, D.E. Browne, T.~Yang, and J.-W. Pan.
\newblock {Demonstration of a Compiled Version of Shor's Quantum Factoring
  Algorithm Using Photonic Qubits}.
\newblock {\em Phys. Rev. Lett.}, 99:250504, 2007.

\bibitem{LGZZYP08}
C.-Y. Lu, W.-B. Gao, J.~Zhang, X.-Q. Zhou, T.~Yang, and J.-W. Pan.
\newblock {Experimental quantum coding against qubit loss error}.
\newblock {\em Proc. Nat. Acad. Sci. USA}, 105:11050, 2008.

\bibitem{L09}
B.P. Lanyon, M.~Barbieri, M.P. Almeida, T.~Jennewein, T.C. Ralph, K.~J. Resch,
  G.J. Pryde, J.L. O'Brien, A.~Gilchrist, and A.G. White.
\newblock {Simplifying quantum logic using higher-dimensional Hilbert spaces}.
\newblock {\em Nature Physics}, 5:134, 2009.

\bibitem{OFV09}
J.L. O'Brien, A.~Furusawa, and J.~Vuckovic.
\newblock {Photonic Quantum Technologies}.
\newblock {\em Nature Photonics}, 3:687, 2009.

\bibitem{PMO09}
A.~Politi, J.C.F. Matthews, and J.L. O'Brien.
\newblock {Shor's quantum factoring algorithm on a photonic chip}.
\newblock {\em Science}, 325:1221, 2009.

\bibitem{MMOY07}
D.L. Moehring, P.~Maunz, S.~Olmschenk, K.C. Younge, D.N. Matsukevich, L.-M.
  Duan, and C.~Monroe.
\newblock {Entanglement of single-atom quantum bits at a distace}.
\newblock {\em Nature (London)}, 449:68, 2007.

\bibitem{OHMMMM10+}
S.~Olmschenk, D.N. Matsukevich, P.~Maunz, D.~Hayes, L.-M. Duan, D.L. Moehring,
  and C.~Monroe.
\newblock {Quantum Teleportation Between Distant Matter Qubits}.
\newblock {\em Science}, 323:486, 2010.

\bibitem{MMKIW95}
C.~Monroe, D.M. Meekhof, B.E. King, W.M. Itano, and D.J. Wineland.
\newblock {Demonstration of a Fundamental quantum logic gate}.
\newblock {\em Phys. Rev. Lett.}, 75:4714, 1995.

\bibitem{CPHS98}
D.G. Cory, M.D. Price, W.~Maas, E.~Knill, R.~Laflamme, W.H. Zurek, T.F. Havel,
  and S.S. Somaroo.
\newblock {Experimental quantum error correction}.
\newblock {\em Phys. Rev. Lett.}, 81:2152, 1998.

\bibitem{SKC00}
C.A. Sackett, D.~Kielpinski, B.E. King, C.~Langer, V.~Meyer, C.J. Myatt,
  M.~Rowe, Q.A. Turchette, W.M. Itano, D.J. Wineland, and C.~Monroe.
\newblock {Experimental entanglement of four particles}.
\newblock {\em Nature (London)}, 404:256, 2000.

\bibitem{SPS07}
M.A. Sillanpaa, J.I. Park, and R.W. Simmonds.
\newblock {Coherent quantum state storage and transfer between two phase qubits
  via a resonant cavity}.
\newblock {\em Nature (London)}, 449:438, 2007.

\bibitem{MCGK07}
J.~Majer, J.M. Chow, J.M. Gambetta, J.~Koch, B.R.Johnson, J.A. Schreier,
  L.~Frunzio, D.I. Schuster, A.A. Houck, A.~Wallraff, A.~Blais, M.H. Devoret,
  S.M. Girvin, and R.J. Schoelkopf.
\newblock {Coupling superconducting qubits via a cavity bus}.
\newblock {\em Nature (London)}, 449:443, 2007.

\bibitem{LBSB10}
Y.~Li, S.D. Barrett, T.M. Stace, and S.C. Benjamin.
\newblock {Fully fault-tolerant quantum computation with non-deterministic
  gates}.
\newblock {\em arxiv:1008.1369}, 2010.

\bibitem{FT10}
K.~Fujii and Y.~Tokunaga.
\newblock {Fault-Tolerant Topological One-Way Quantum Computation with
  Probabilistic Two-Qubit Gates}.
\newblock {\em Phys. Rev. Lett.}, 105:250503, 2010.

\bibitem{K97}
A.Y. Kitaev.
\newblock {Quantum Computations: algorithms and error correction}.
\newblock {\em Russ. Math. Serv.}, 52(6):1191, 1997.

\bibitem{SSRG09}
N.~Sangouard, C.~Simon, H.~de~Riedmatten, and N.~Gisin.
\newblock {Quantum repeaters based on atomic ensembles and linear optics}.
\newblock {\em arxiv:0906.2699}, 2009.

\bibitem{SGMNH08}
C-H. Su, A.D. Greentree, W.J. Munro, K.~Nemoto, and L.C.L. Hollenberg.
\newblock {High Speed quantum gates with cavity quantum electrodynamics}.
\newblock {\em Phys. Rev. A.}, 78:062336, 2008.

\bibitem{NM05}
K.~Nemoto and W.J. Munro.
\newblock {Universal quantum computation on the power of quantum non-demolition
  measurements}.
\newblock {\em Phys. Lett. A}, 344:104, 2005.

\bibitem{MNS05}
W.J. Munro, K.~Nemoto, and T.P. Spiller.
\newblock {Weak nonlinearities: a new route to optical quantum computation}.
\newblock {\em New. J. Phys.}, 7:137, 2005.

\bibitem{BS10}
S.D. Barrett and T.M. Stace.
\newblock {Fault-Tolerant quantum computation with very high threshold for loss
  errors}.
\newblock {\em Phys. Rev. Lett.}, 105:200502, 2010.

\end{thebibliography}

\appendix
\section{Further Simulations: Number of photons as a function of timesteps}
\label{sec:AppA}

Shown here are further simulations for $8 \leq N \leq 168$, showing the total number of photons in the 
network as a function of total timesteps, $t$, and bias, $B$.  Total samples are $10^3$ and for larger networks 
the maximum bias simulated is reduced due to computational resources.
\begin{figure}
\begin{center}
\resizebox{85mm}{!}{\includegraphics{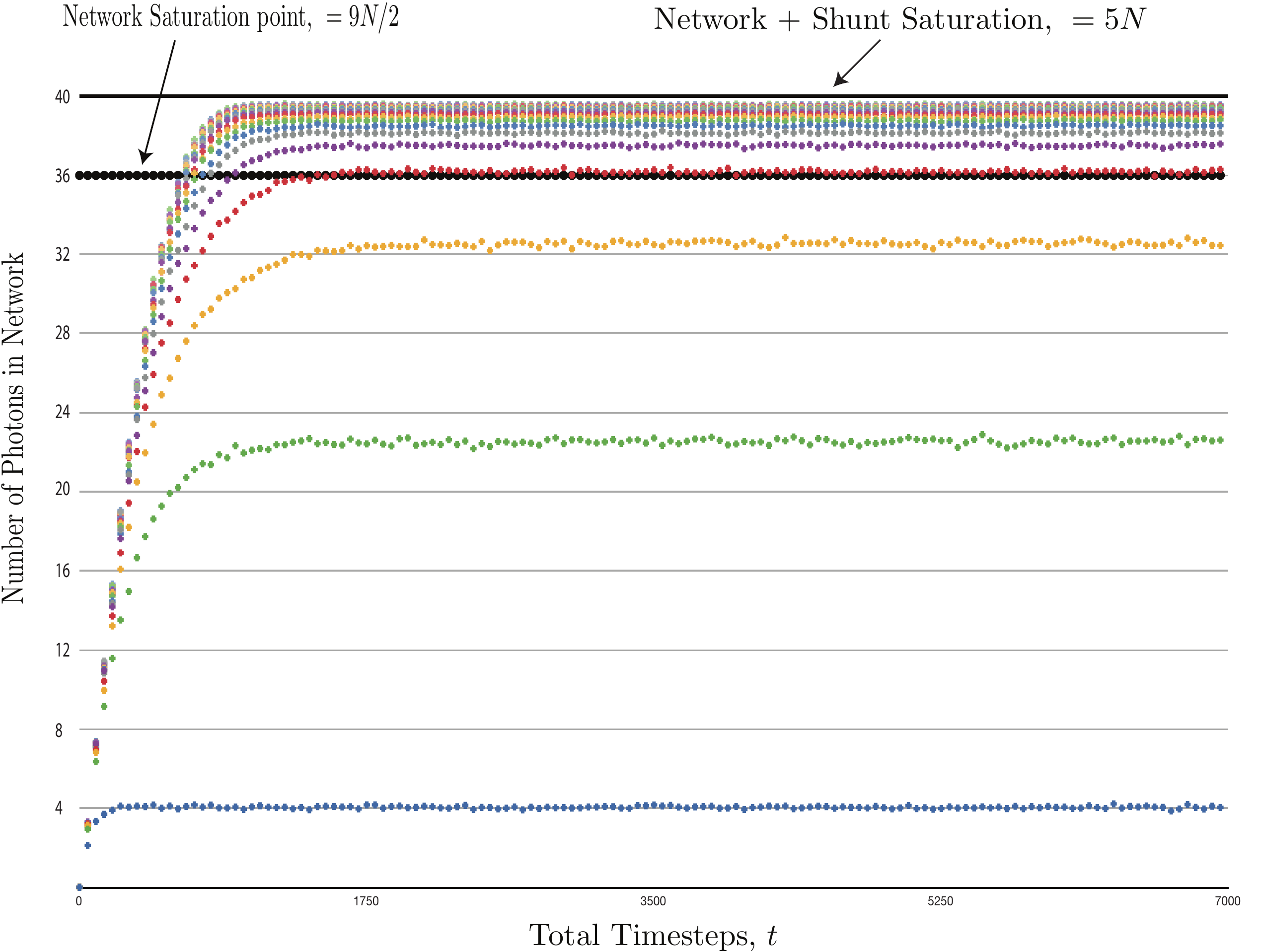}}
\end{center}
\vspace*{-10pt}
\caption{$N=8$. Simulations of $2\leq B \leq 192$.  As the total number of photons is quite low, this simulation suffers from more statistical 
variance for $10^3$ samples.}
\end{figure}
\begin{figure}
\begin{center}
\resizebox{85mm}{!}{\includegraphics{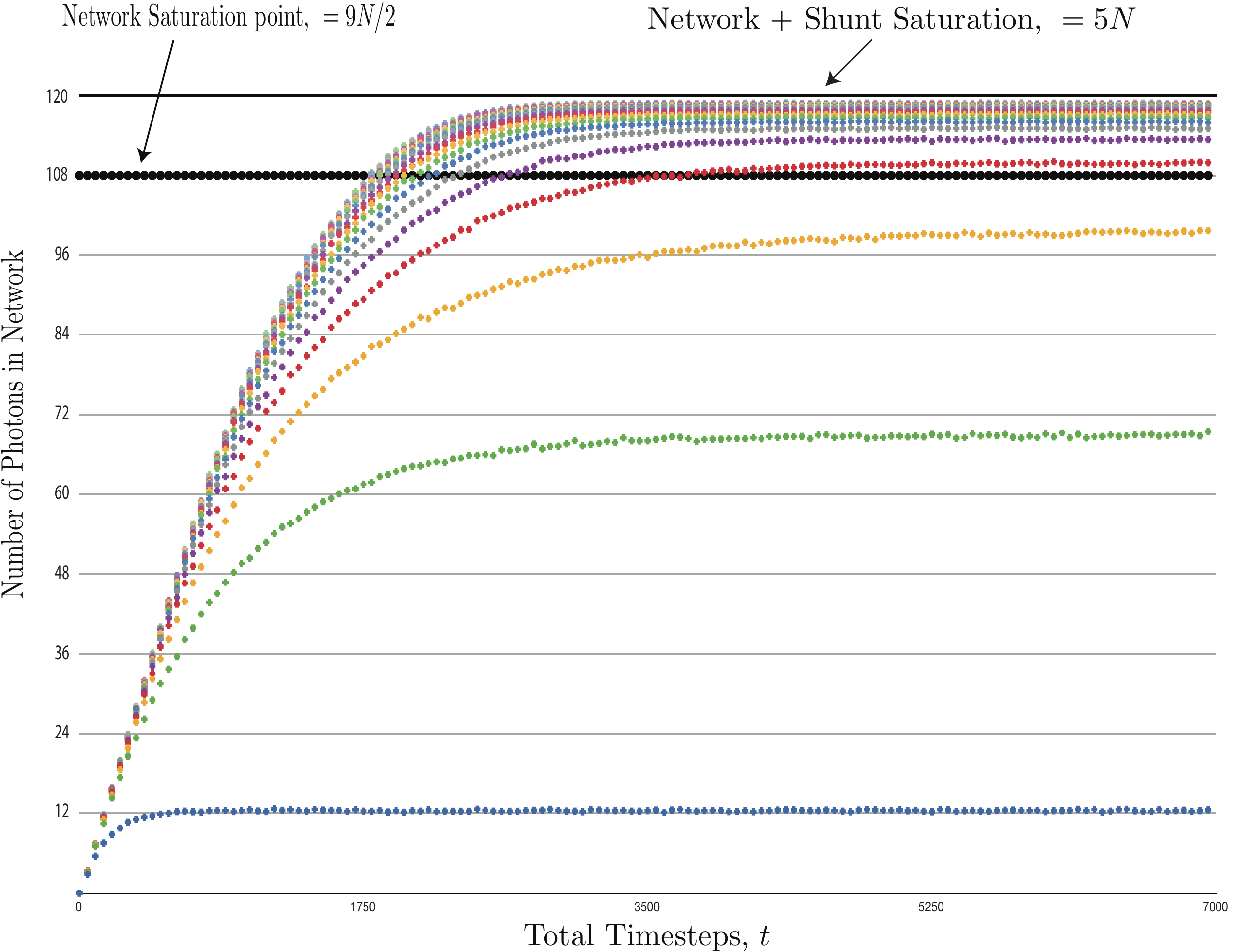}}
\end{center}
\vspace*{-10pt}
\caption{$N=24$. Simulations of $2\leq B \leq 192$.}
\end{figure}
\begin{figure}
\begin{center}
\resizebox{85mm}{!}{\includegraphics{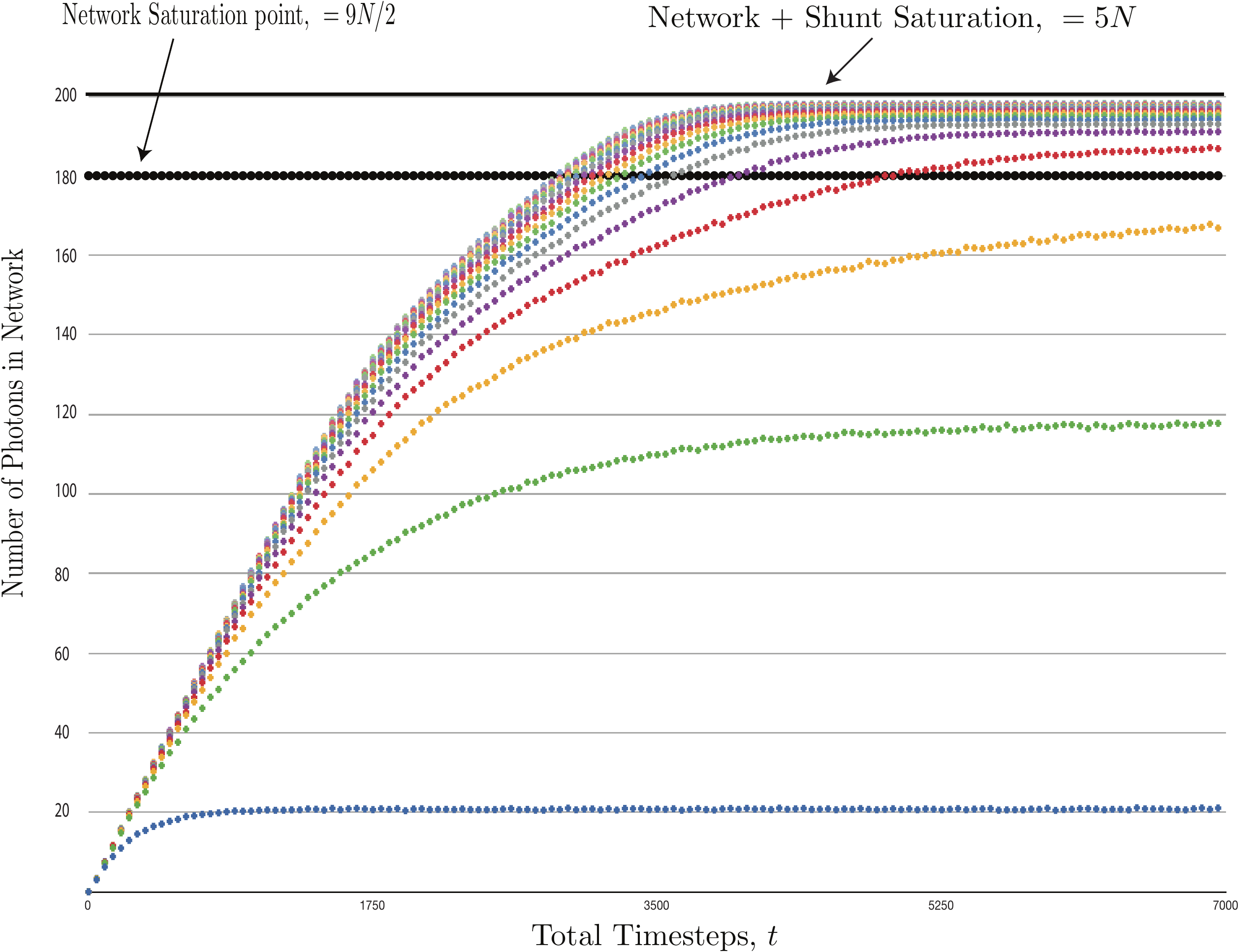}}
\end{center}
\vspace*{-10pt}
\caption{$N=40$. Simulations of $2\leq B \leq 192$.}
\end{figure}
\begin{figure}
\begin{center}
\resizebox{85mm}{!}{\includegraphics{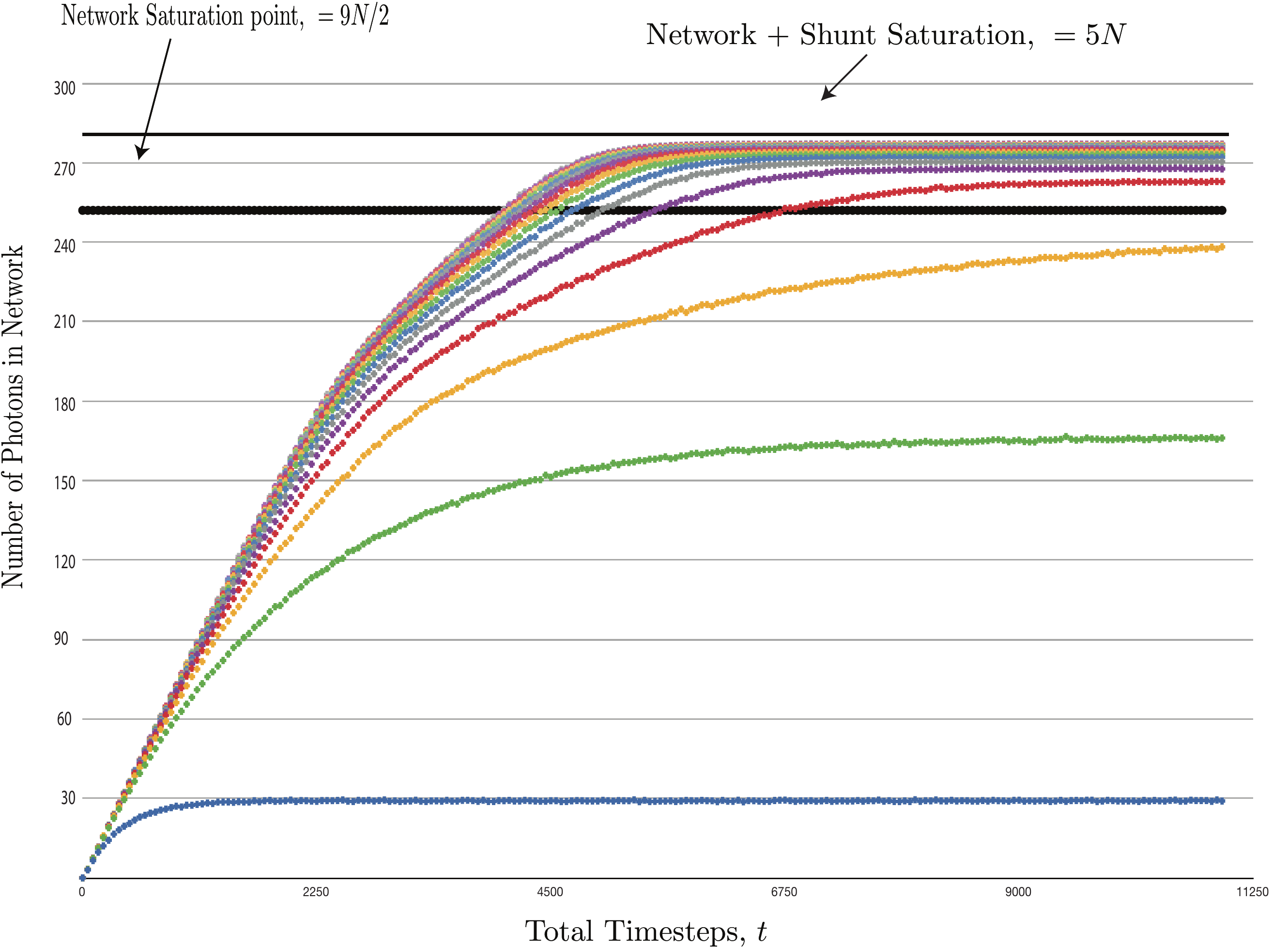}}
\end{center}
\vspace*{-10pt}
\caption{$N=56$. Due to computational resources, we restrict simulations to $2\leq B \leq 172$.}
\end{figure}
\begin{figure}
\begin{center}
\resizebox{85mm}{!}{\includegraphics{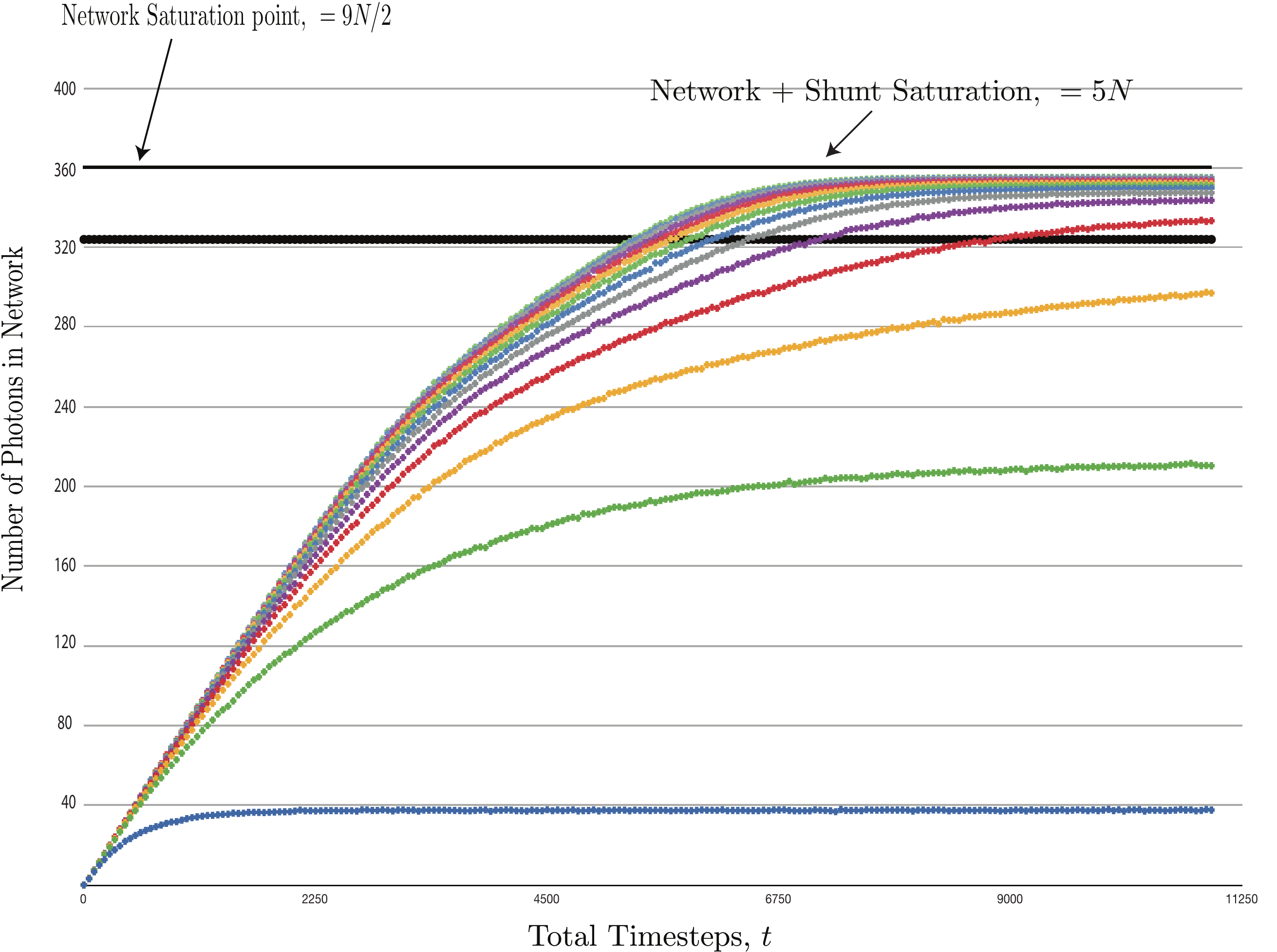}}
\end{center}
\vspace*{-10pt}
\caption{$N=72$. Due to computational resources, we restrict simulations to $2\leq B \leq 132$.}
\end{figure}
\begin{figure}
\begin{center}
\resizebox{85mm}{!}{\includegraphics{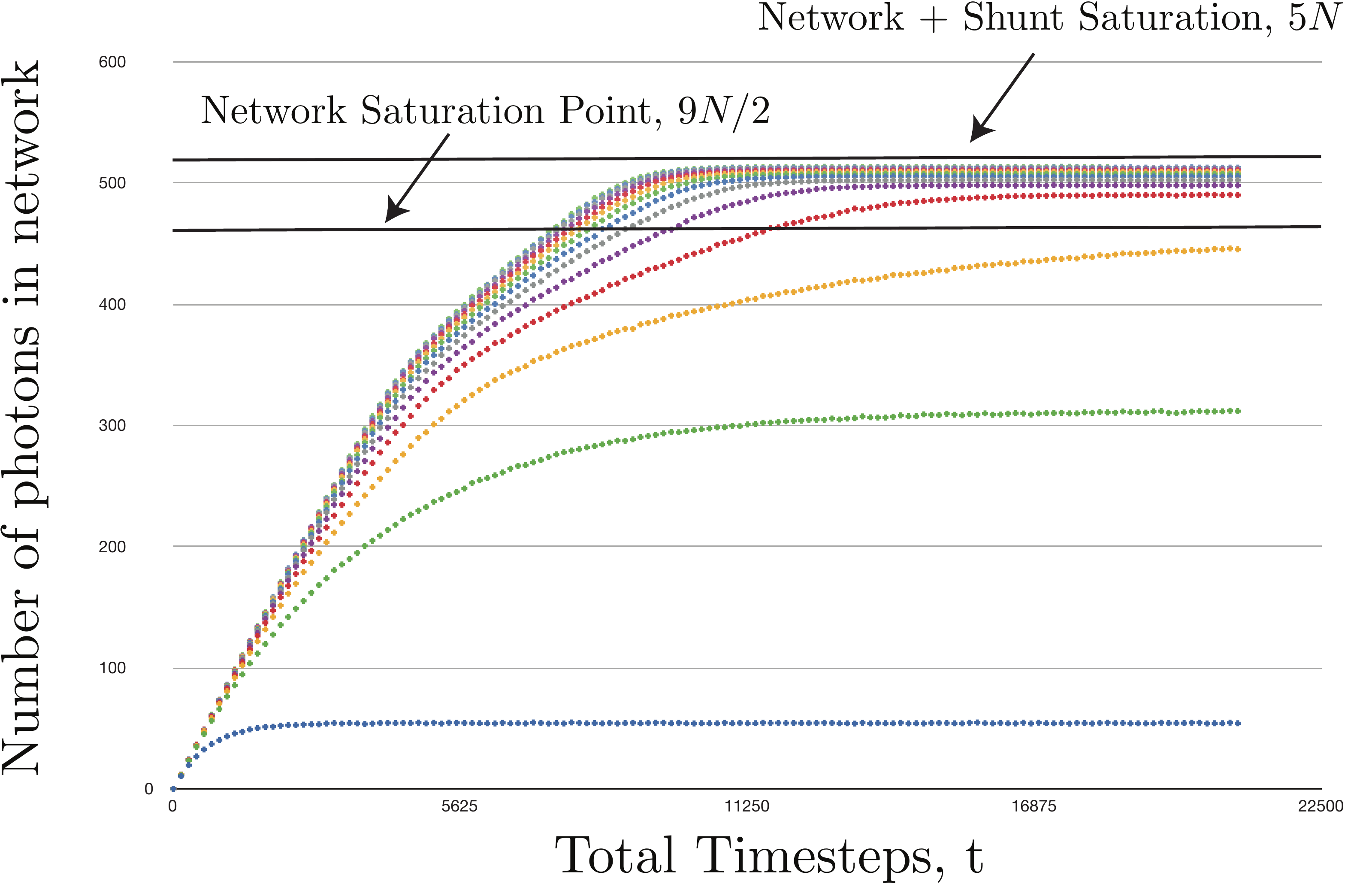}}
\end{center}
\vspace*{-10pt}
\caption{$N=104$. Due to computational resources, we restrict simulations to $2\leq B \leq 132$.}
\end{figure}
\begin{figure}[ht!]
\begin{center}
\resizebox{85mm}{!}{\includegraphics{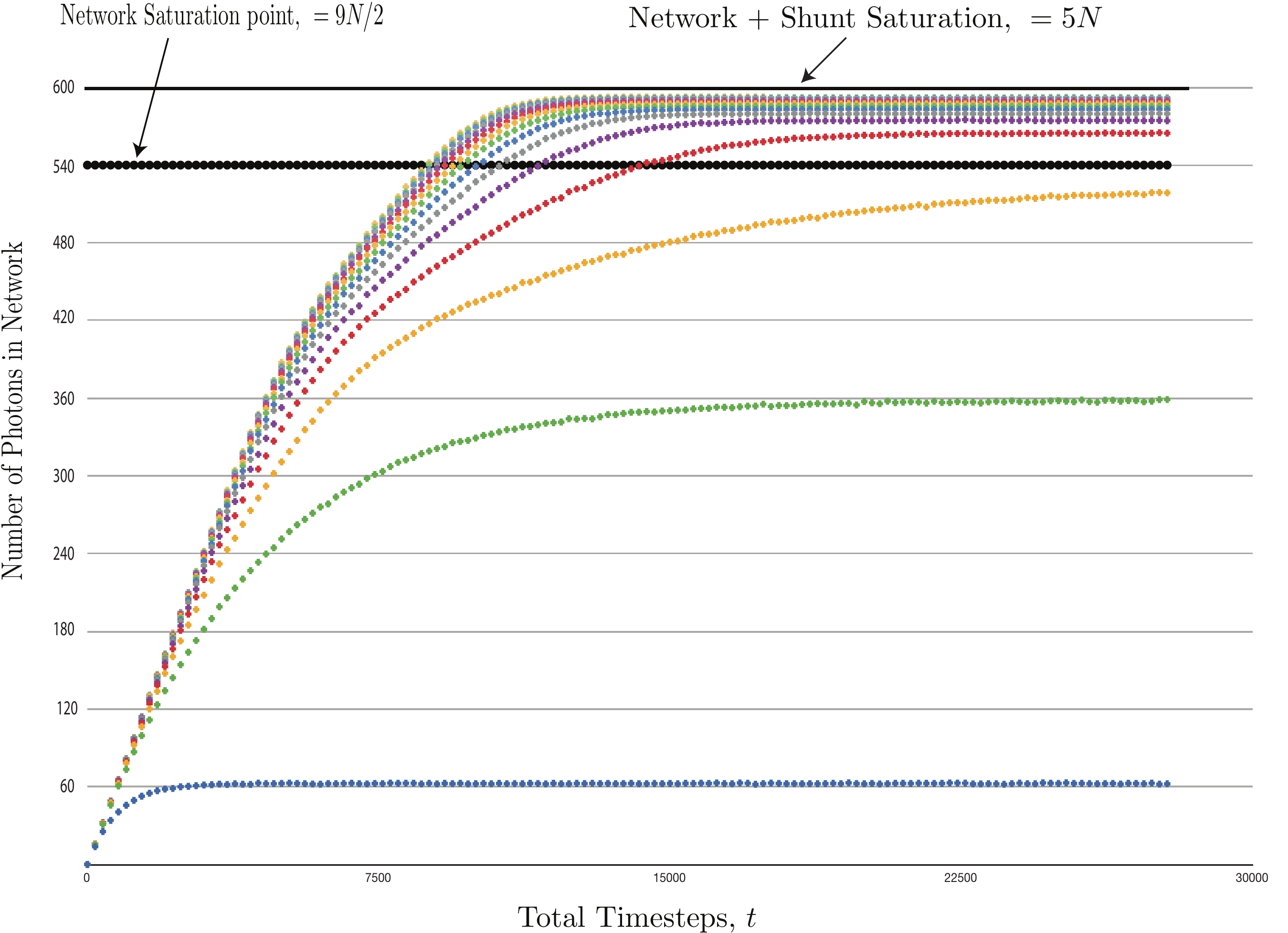}}
\end{center}
\vspace*{-10pt}
\caption{$N=120$.  Due to computational resources, we restrict simulations to $2\leq B \leq 132$.}
\end{figure}
\begin{figure}
\begin{center}
\resizebox{85mm}{!}{\includegraphics{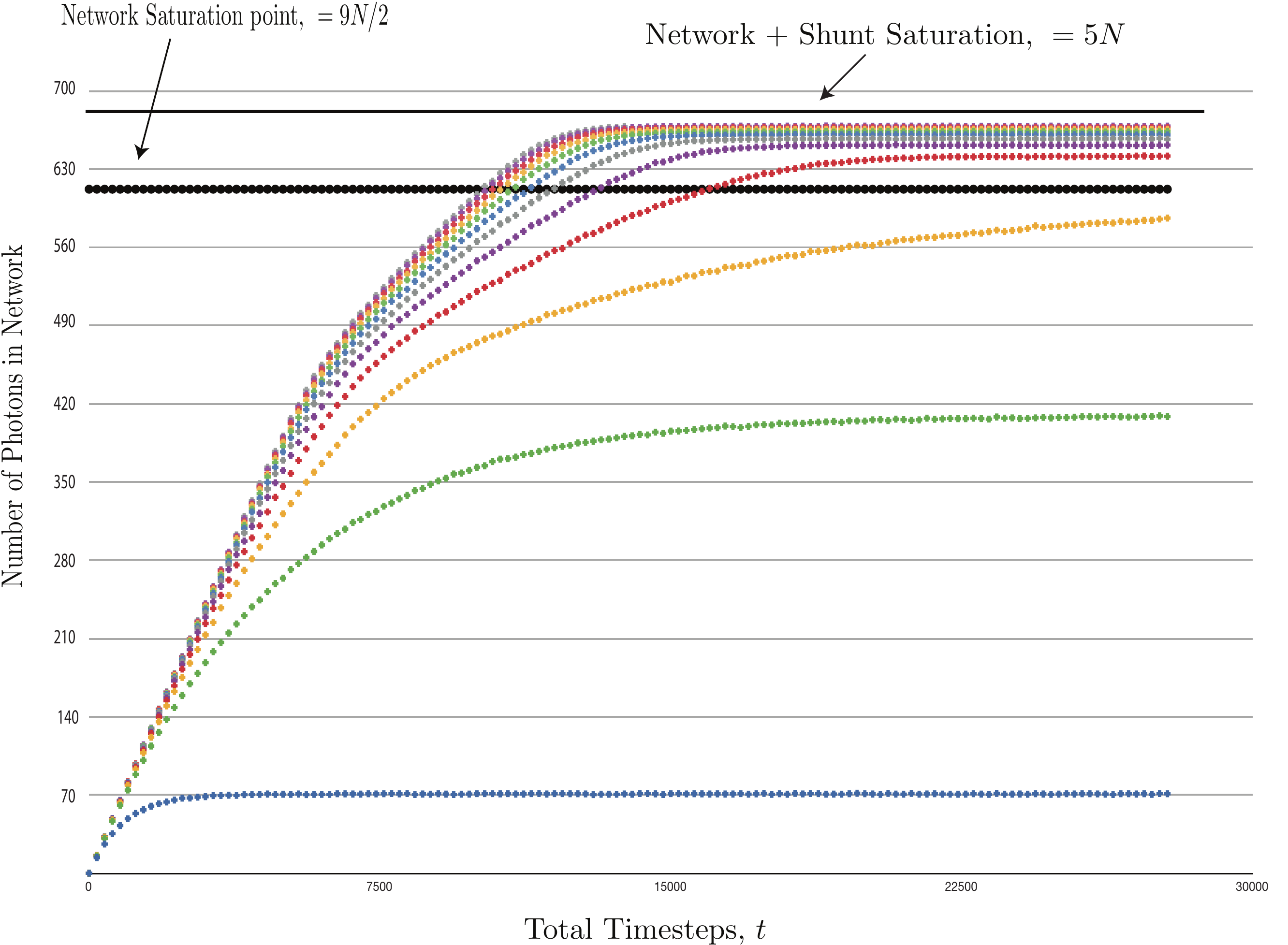}}
\end{center}
\vspace*{-10pt}
\caption{$N=136$.  Due to computational resources, we restrict simulations to $2\leq B \leq 102$.}
\end{figure}
\begin{figure}
\begin{center}
\resizebox{85mm}{!}{\includegraphics{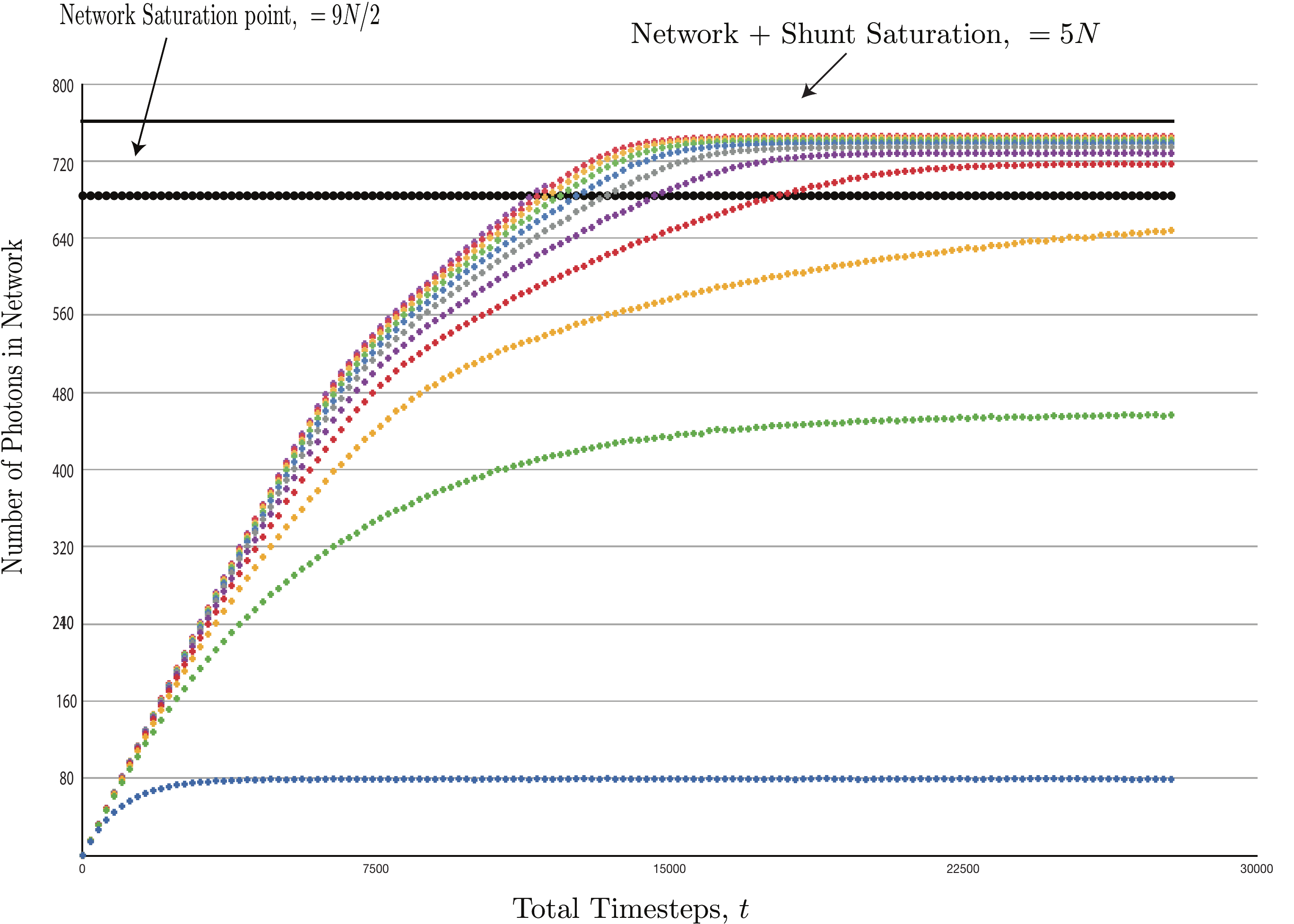}}
\end{center}
\vspace*{-10pt}
\caption{$N=152$.  Due to computational resources, we restrict simulations to $2\leq B \leq 92$.}
\end{figure}
\begin{figure}
\begin{center}
\resizebox{85mm}{!}{\includegraphics{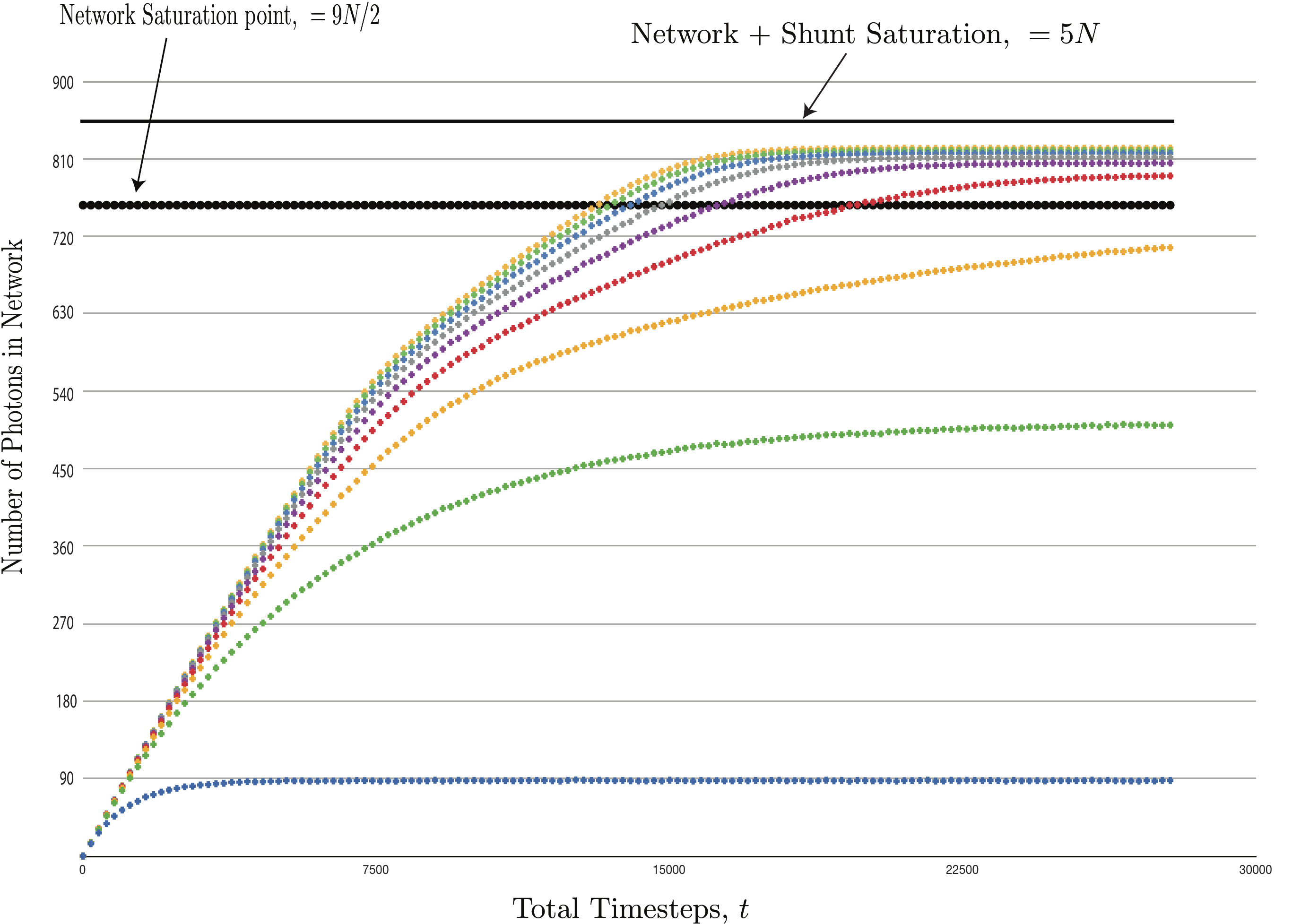}}
\end{center}
\vspace*{-10pt}
\caption{$N=168$.  Due to computational resources, we restrict simulations to $2\leq B \leq 82$.}
\end{figure}

\section{Further Simulations: Number of photons as a function of Bias}
\label{sec:AppB}

Shown here are further simulations for $8 \leq N \leq 168$, showing the total number of photons in the 
network as a function of bias, $B$.  Each data point is taken at the maximum value of $t$ simulated in 
Appendix. \ref{sec:AppA}. For each network size, the system saturates at $B\approx 30$.  The threshold 
bias decreases slightly as $N$ increases.  

\begin{figure}
\begin{center}
\resizebox{85mm}{!}{\includegraphics{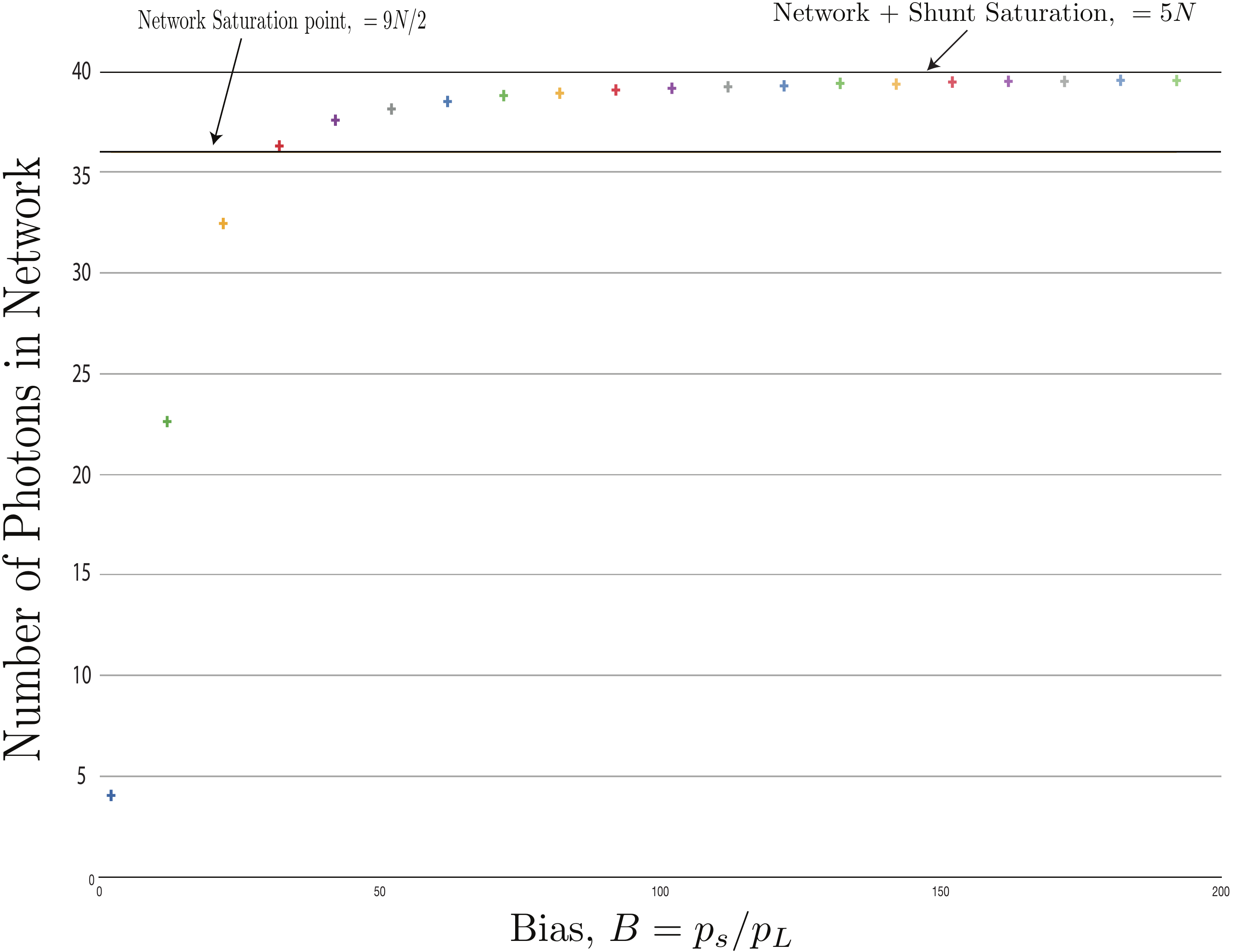}}
\end{center}
\vspace*{-10pt}
\caption{$N=8$. Simulations of $2\leq B \leq 192$.}
\end{figure}
\begin{figure}
\begin{center}
\resizebox{85mm}{!}{\includegraphics{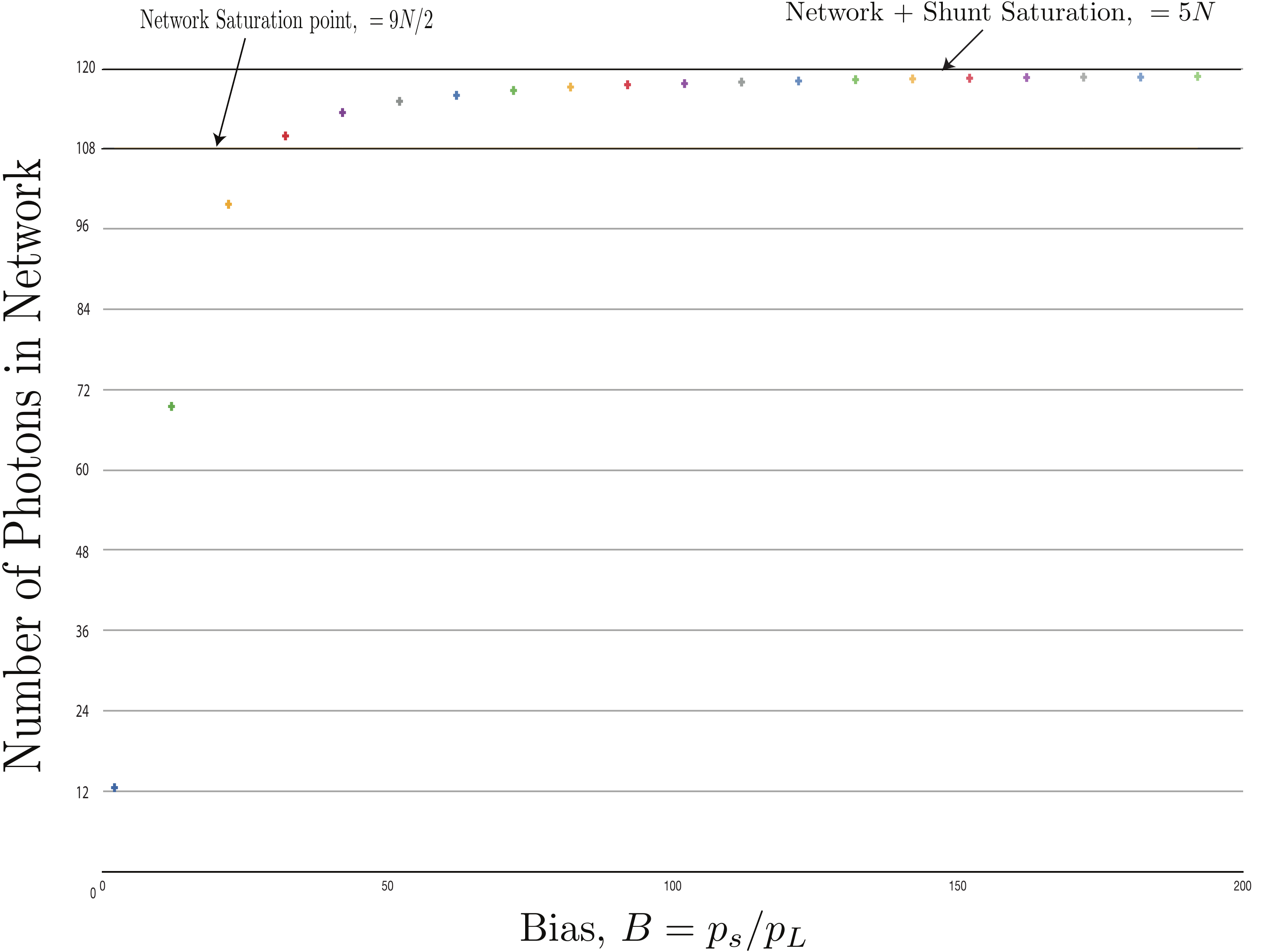}}
\end{center}
\vspace*{-10pt}
\caption{$N=24$. Simulations of $2\leq B \leq 192$.}
\end{figure}
\begin{figure}
\begin{center}
\resizebox{85mm}{!}{\includegraphics{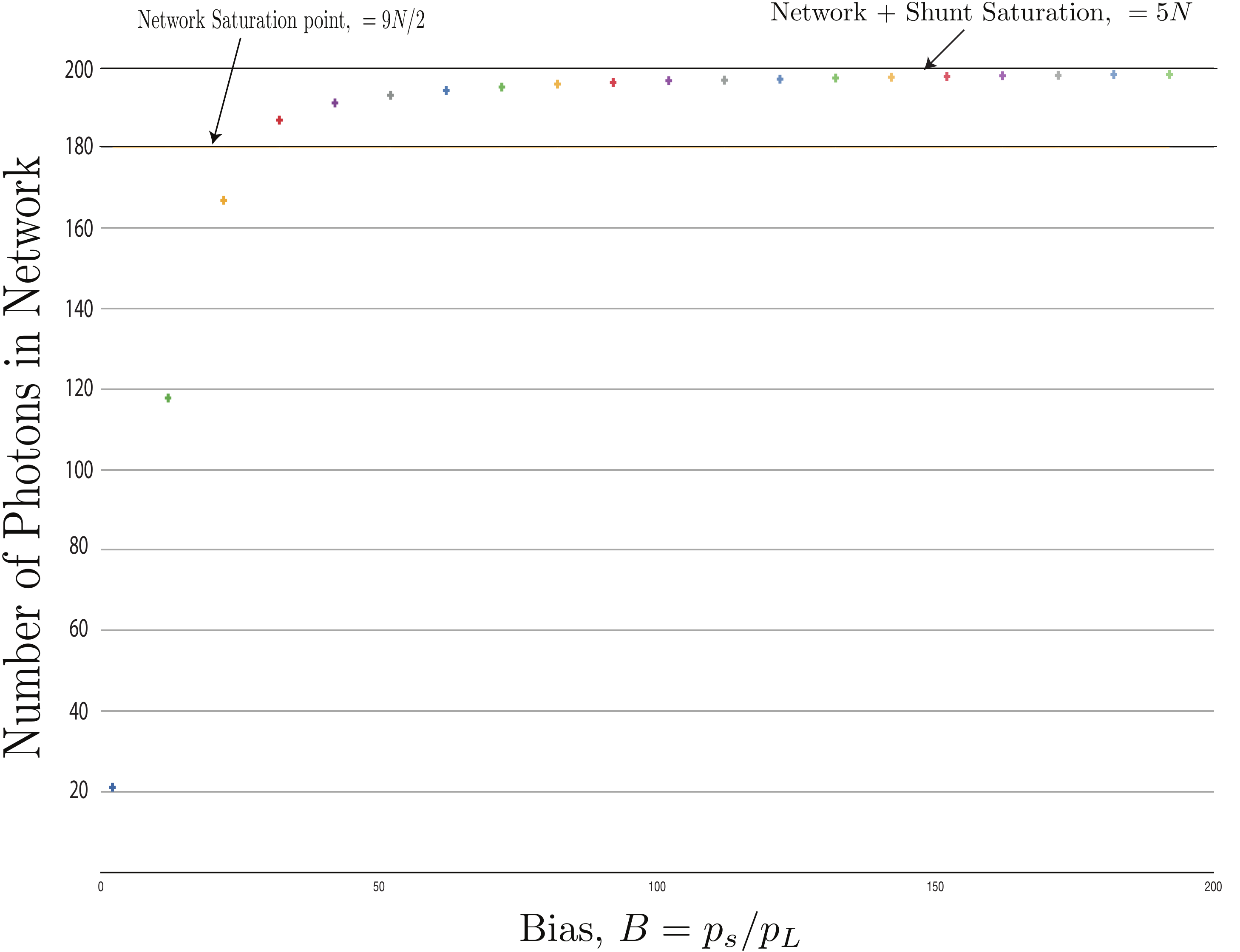}}
\end{center}
\vspace*{-10pt}
\caption{$N=40$. Simulations of $2\leq B \leq 192$.}
\end{figure}
\begin{figure}
\begin{center}
\resizebox{85mm}{!}{\includegraphics{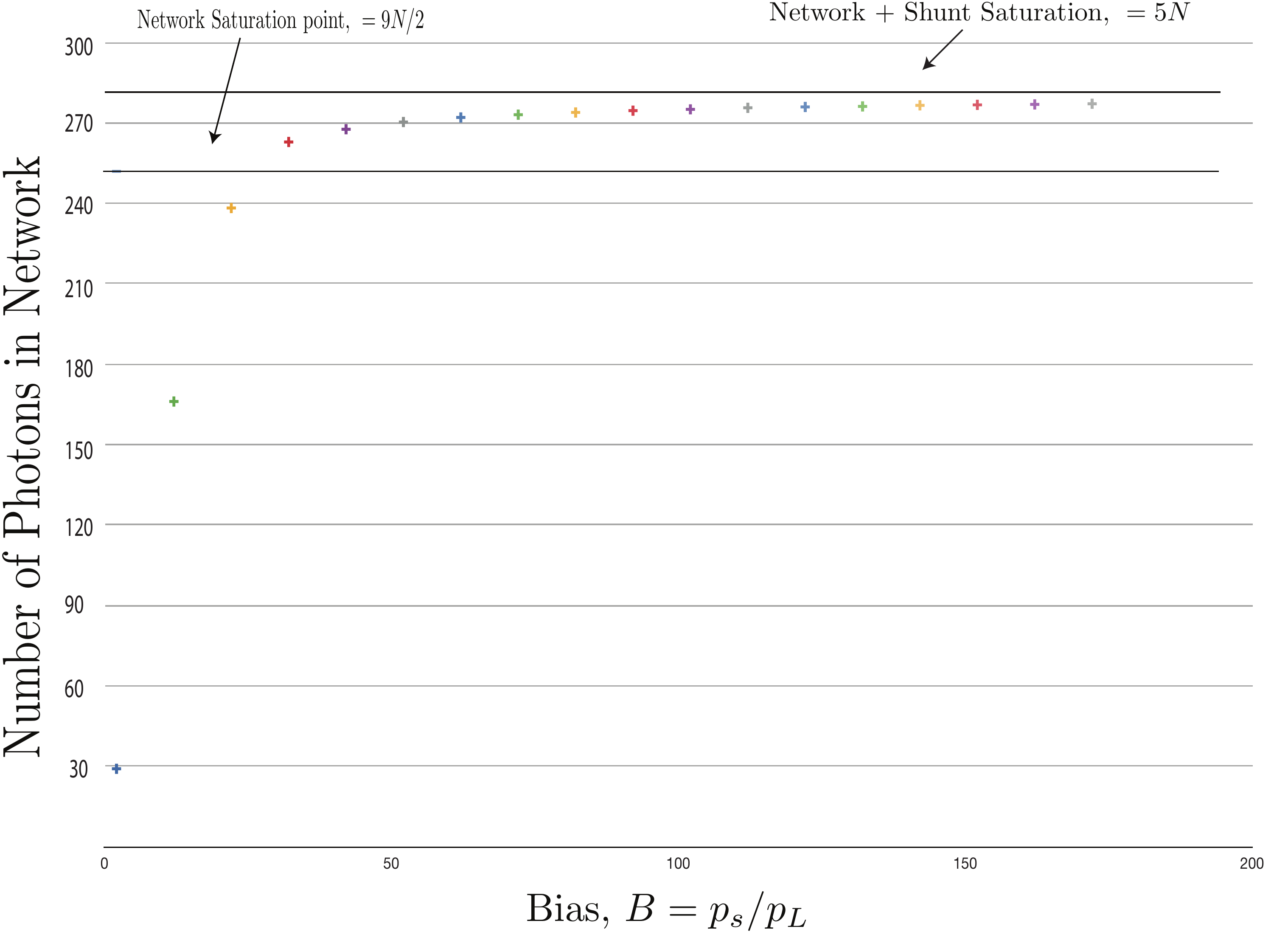}}
\end{center}
\vspace*{-10pt}
\caption{$N=56$.  Due to computational resources, we restrict simulations to $2\leq B \leq 172$.}
\end{figure}
\begin{figure}
\begin{center}
\resizebox{85mm}{!}{\includegraphics{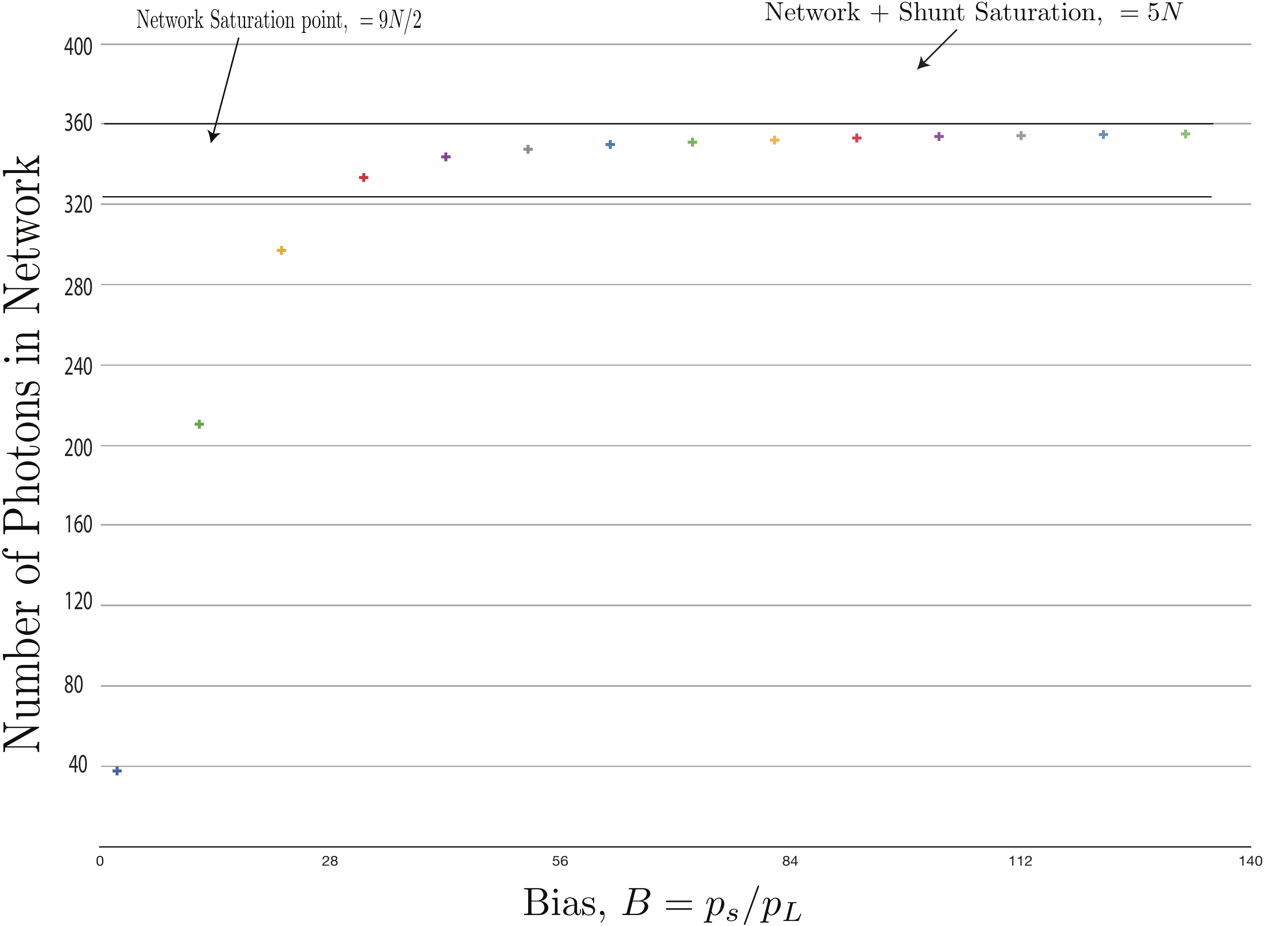}}
\end{center}
\vspace*{-10pt}
\caption{$N=72$.  Due to computational resources, we restrict simulations to $2\leq B \leq 132$.}
\end{figure}
\begin{figure}
\begin{center}
\resizebox{85mm}{!}{\includegraphics{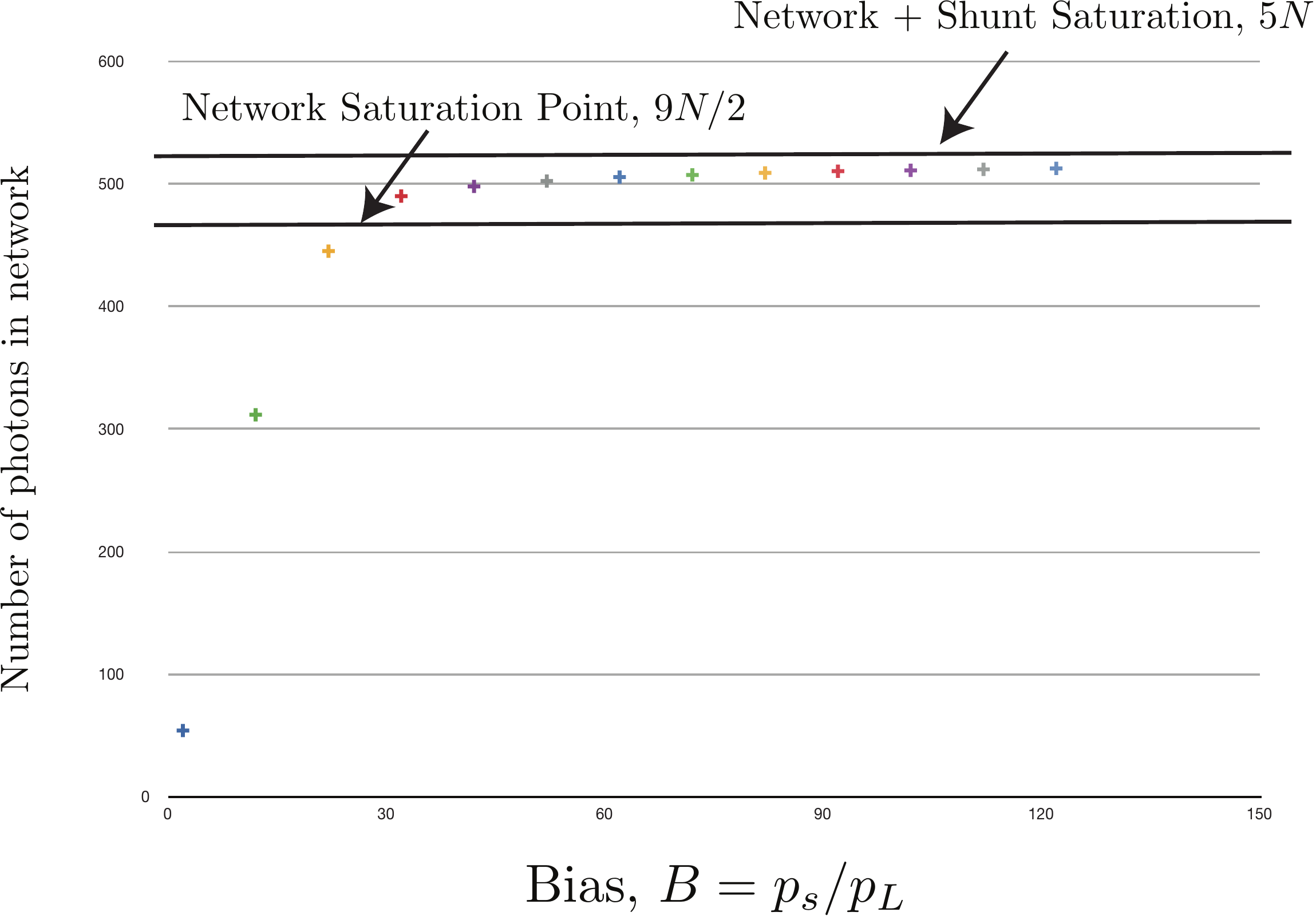}}
\end{center}
\vspace*{-10pt}
\caption{$N=104$. Due to computational resources, we restrict simulations to $2\leq B \leq 132$.}
\end{figure}
\begin{figure}
\begin{center}
\resizebox{85mm}{!}{\includegraphics{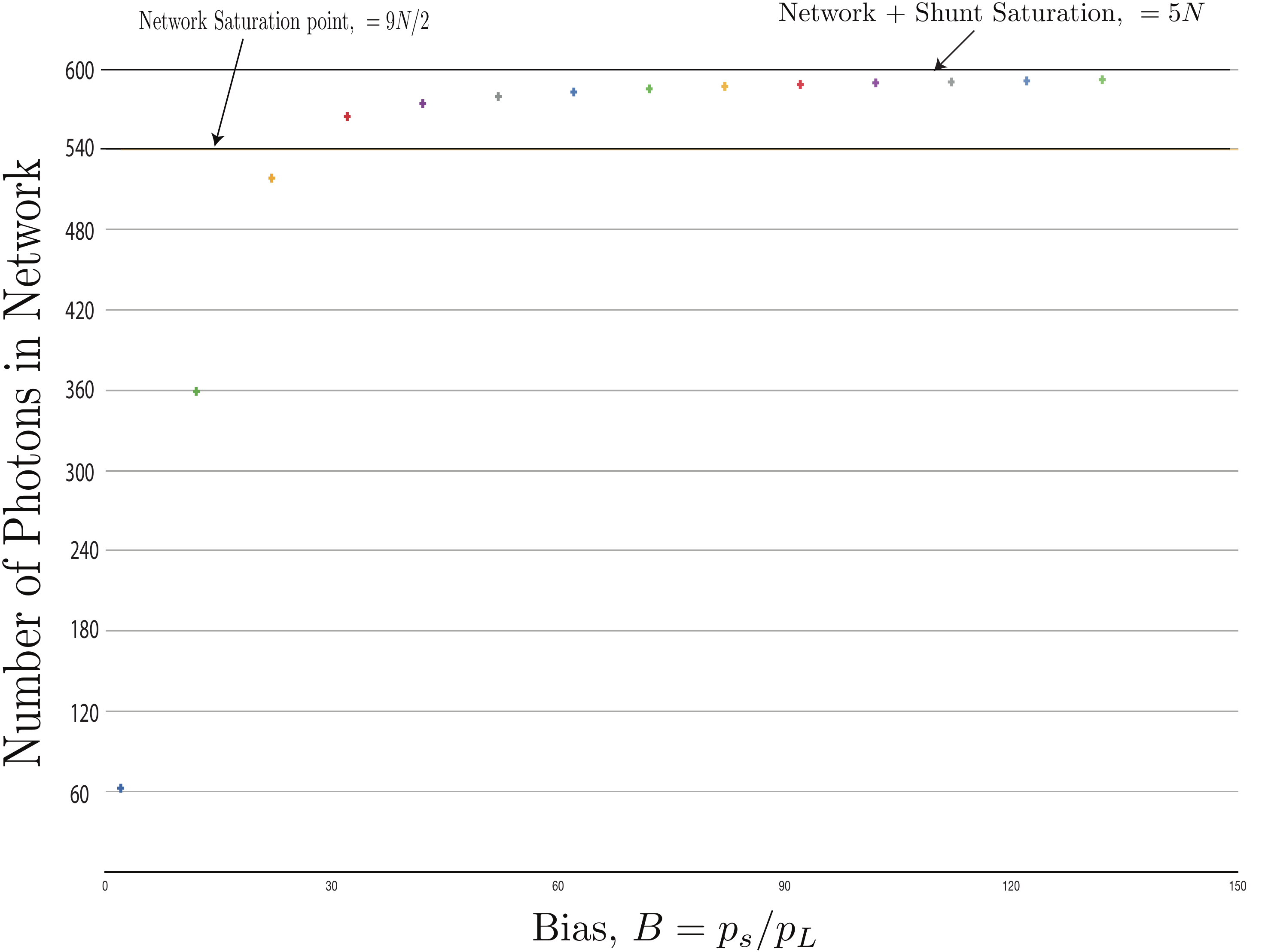}}
\end{center}
\vspace*{-10pt}
\caption{$N=120$.  Due to computational resources, we restrict simulations to $2\leq B \leq 132$.}
\end{figure}
\begin{figure}
\begin{center}
\resizebox{85mm}{!}{\includegraphics{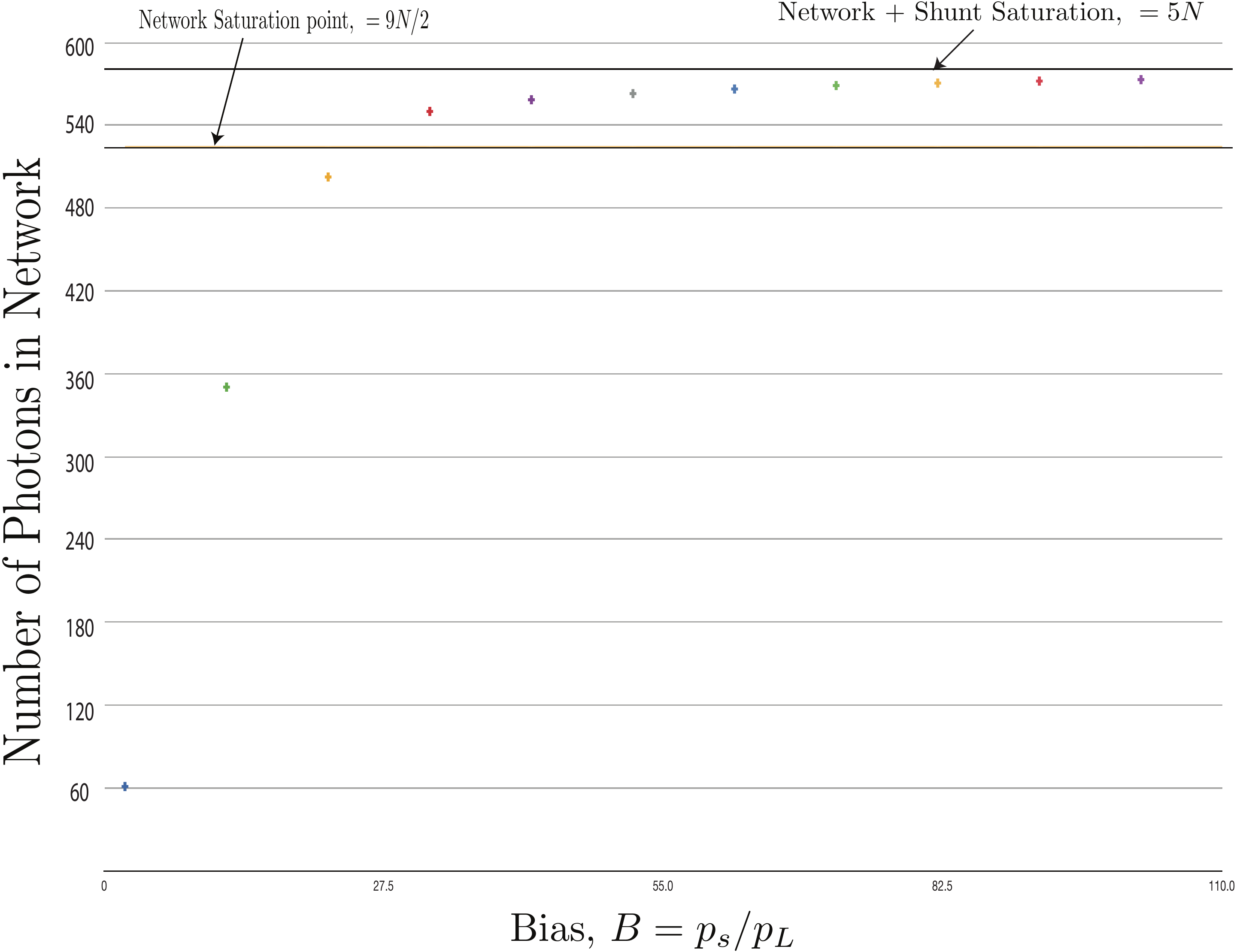}}
\end{center}
\vspace*{-10pt}
\caption{$N=136$.  Due to computational resources, we restrict simulations to $2\leq B \leq 102$.}
\end{figure}
\begin{figure}
\begin{center}
\resizebox{85mm}{!}{\includegraphics{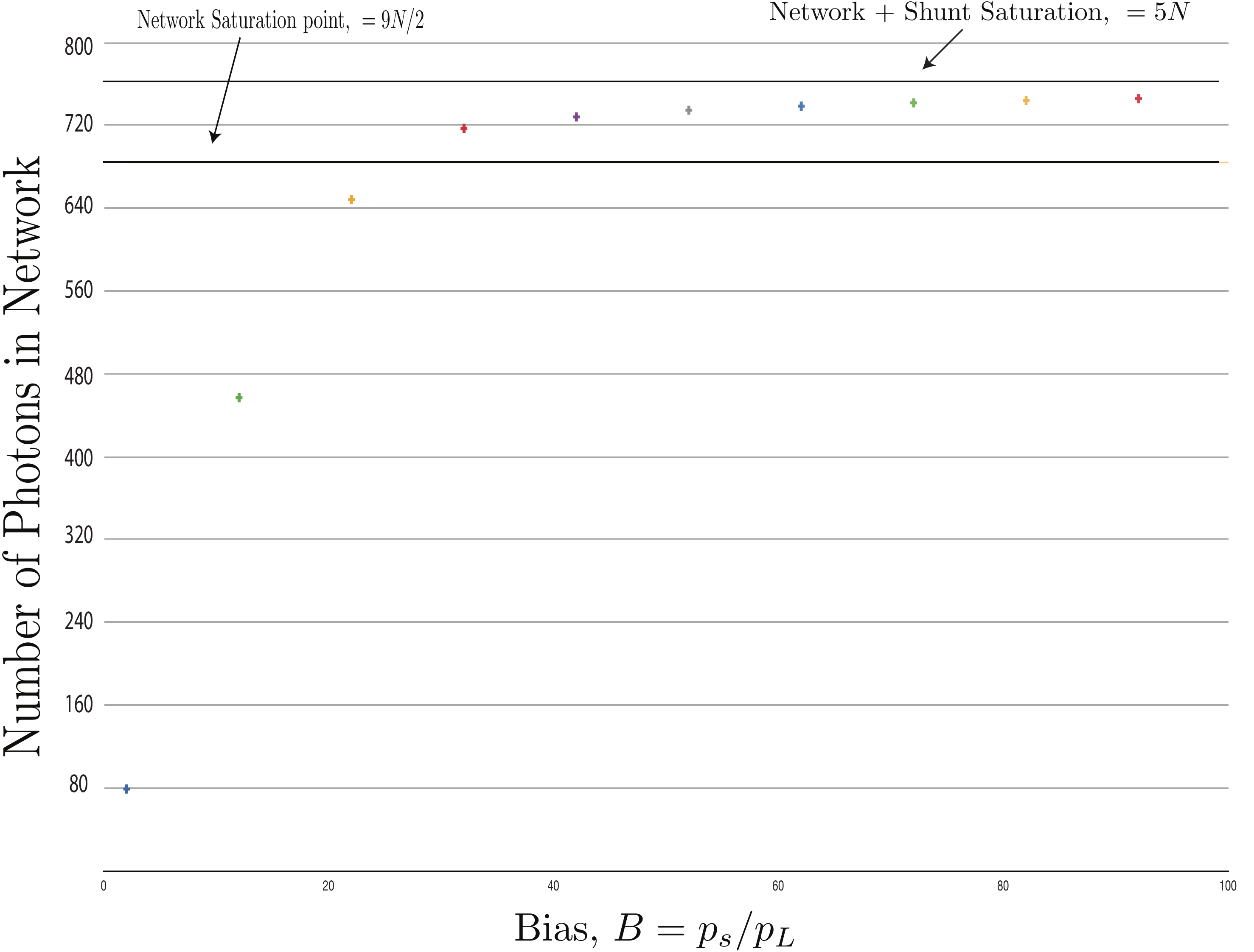}}
\end{center}
\vspace*{-10pt}
\caption{$N=152$.  Due to computational resources, we restrict simulations to $2\leq B \leq 92$.}
\end{figure}
\begin{figure}
\begin{center}
\resizebox{85mm}{!}{\includegraphics{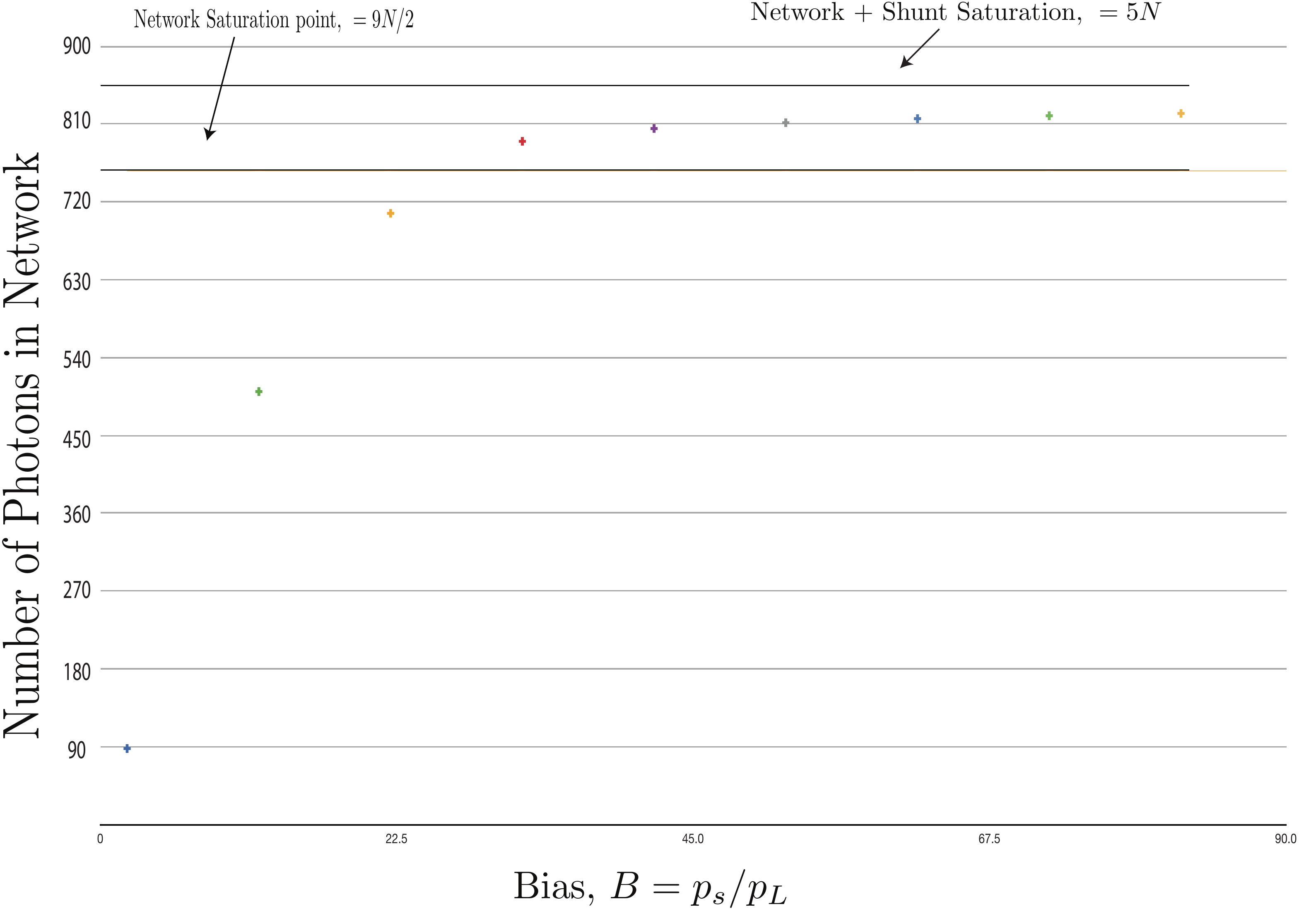}}
\end{center}
\vspace*{-10pt}
\caption{$N=168$.  Due to computational resources, we restrict simulations to $2\leq B \leq 82$.}
\end{figure}

\end{document}